\documentclass[acmsmall]{acmart}
\usepackage{tabularx}
\usepackage{multirow}
\usepackage{float}
\usepackage{subfig}
\usepackage{tikz}
\usepackage[normalem]{ulem}

\AtBeginDocument{%
  \providecommand\BibTeX{{%
    \normalfont B\kern-0.5em{\scshape i\kern-0.25em b}\kern-0.8em\TeX}}}

\usepackage{amsmath,amsfonts}
\usepackage{algorithm}
\usepackage{algorithmicx}
\usepackage{algpseudocode}

\usepackage{graphicx}
\usepackage{textcomp}
\usepackage{xcolor}
\usepackage{multirow} 
\usepackage{float}
\usepackage{subfig}
\newcommand{\ours}{\textsc{Tao}}
\usepackage{hyperref}       
\usepackage{url}            
\usepackage{booktabs}       

\usepackage{nicefrac}       
\usepackage{microtype}      
\usepackage{wrapfig}

\newcommand{\san}[1]{{\color{black} {#1}}}
\newcommand{\update}[1]{{\color{black}{#1}}}

\setcopyright{acmlicensed}
\acmJournal{POMACS}
\acmYear{2024} \acmVolume{8} \acmNumber{2} \acmArticle{28} \acmMonth{6}\acmDOI{10.1145/3656012}

\begin{document}

\title{{\ours}: Re-Thinking DL-based Microarchitecture Simulation}

\author{Santosh Pandey}
\email{santosh.pandey@rutgers.edu}
\affiliation{%
 \institution{Rutgers University}
 \city{New Brunswick}
 \state{NJ}
 \country{USA}
}

\author{Amir Yazdanbakhsh}
\email{ayazdan@google.com}
\affiliation{%
 \institution{Google DeepMind}
 \city{Mountain View}
 \state{CA}
 \country{USA}
}

\author{Hang Liu}
\email{hl1097@soe.rutgers.edu}
\affiliation{%
 \institution{Rutgers University}
 \city{New Brunswick}
 \state{NJ}
 \country{USA}
}

\begin{abstract}

Microarchitecture simulators are indispensable tools for microarchitecture designers to validate, estimate, optimize, and manufacture new hardware that meets specific design requirements. While the quest for a fast, accurate and detailed microarchitecture simulation has been ongoing for decades, existing simulators excel and fall short at different aspects: (i) Although execution-driven simulation is accurate and detailed, it is extremely slow and requires expert-level experience to design. (ii) Trace-driven simulation reuses the execution traces in pursuit of fast simulation but faces accuracy concerns and fails to achieve significant speedup. (iii) Emerging deep learning (DL)-based simulations are remarkably fast and have acceptable accuracy, but fail to provide adequate low-level microarchitectural performance metrics such as branch mispredictions or cache misses, which is crucial for microarchitectural bottleneck analysis.
Additionally, they introduce substantial overheads from trace regeneration and model re-training when simulating a new microarchitecture. 

Re-thinking the advantages and limitations of the aforementioned three mainstream simulation paradigms, this paper introduces {\ours} that redesigns the DL-based simulation with three primary contributions: First, we propose a new training dataset design such that the subsequent simulation (i.e., inference) only needs functional trace as inputs, which can be rapidly generated and reused across microarchitectures. Second, to increase the detail of the simulation, we redesign the input features and the DL model using self-attention to support predicting various performance metrics of interest. Third, we propose techniques to train a microarchitecture agnostic embedding layer that enables fast transfer learning between different microarchitectural configurations and effectively reduces the re-training overhead of conventional DL-based simulators.
{\ours} can predict various performance metrics of interest, significantly reduce the simulation time, and maintain similar simulation accuracy as state-of-the-art DL-based endeavors. Our extensive evaluation shows {\ours} can reduce the overall training and simulation time  
by 18.06$\times$ over the state-of-the-art DL-based endeavors.


\end{abstract}



\maketitle 

\section{Introduction}


Since its inception, microarchitecture simulators rapidly become the most commonly used tools in computer architecture-related research (see the report~\cite{skadron2003challenges}). As of today, computer architecture simulation is the textbook standard and virtually used in any architecture explorations, e.g., design space exploration~\cite{joseph2006construction,kim2018machine,yazdanbakhsh2021apollo,krishnan2023archgym}, microarchitectural bottleneck analysis~\cite{fields2003using, bai2023archexplorer}, workload characterization~\cite{hoste2007microarchitecture,najaf2008configurational} among many others~\cite{lim2014power,ye2000design}. As a common practice, architecture researchers often use popular software architecture simulators to incorporate their radical new ideas. The updated simulator is then used to execute the programs of interest (i.e., benchmarks). The simulation yields a range of metrics that characterize the execution of benchmarks, with the level of detail in the simulation dictating the specificity of these metrics. Such output metrics provide feedback to the researcher for further explorations and/or decision makings. Due to the significance, many simulators have been built over the decades with different abstractions where each abstraction provides a tradeoff between speed, accuracy and detail (please see~\cite{brais2020survey,uhlig1997trace, carlson2011sniper, akram2019survey} for more details). 

 \begin{figure}
	\centering
	\includegraphics[width=\linewidth]{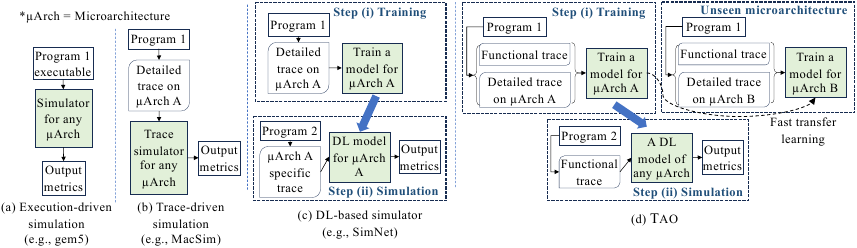}	
    \vspace{-0.2in}
    \caption{\update{Mainstream simulation mechanisms vs. our effort, i.e., {\ours}}.}   
    \vspace{-0.2in}
    \label{fig:introduction}
\end{figure}

\subsection{Related work and motivations}
\label{sec:intro:related}


The quest towards a \textit{fast}, \textit{accurate} and \textit{detailed} cycle-level architecture simulation has never stopped. This cohort of researchers have mainly dedicated their efforts into three prominent paradigms, i.e., execution-driven simulation~\cite{binkert2011gem5,ortego2004sesc,austin2002simplescalar,ye2000design, kanev2012xiosim,ahn2013mcsima+,miller2010graphite,sanchez2013zsim}, trace-driven simulation~\cite{cmelik1994shade,kim2012macsim,alomar2023causalsim, rico2011trace,arafa2021hybrid, khairy2020accel}, and recently the DL-based simulation~\cite{simnet,mendis2019ithemal,sykora2022granite,pandey2022scalable} (see~\cite{brais2020survey,carlson2011sniper, ubal2012multi2sim, bellard2005qemu} for more types of architecture simulations).
Below, we briefly discuss these simulation methodologies with the motivations for our work, i.e., {\ours}.


%
\textbf{Motivation (i).} \san{Execution-driven simulation offers the most detailed and accurate framework, although this comes at the cost of extremely slow speed and high maintenance overhead. }
%
%
%
Figure~\ref{fig:introduction}(a) presents the workflow of this paradigm. 
Takes as input of an executable for a program 1, this simulation approach uses software components to simulate the functionality and timing information of all components of \san{a target} 
processor.
The output statistics contain the runtime behaviors of the hardware when the executable runs through the simulator, including CPI, branch mispredictions, cache misses and many other performance metrics. \san{These performance metrics aid in microarchitectural bottleneck analysis and hardware design exploration.}
The milestone projects of this line of efforts encompass SimpleScalar~\cite{austin2002simplescalar}, SESC~\cite{ortego2004sesc}, and gem5~\cite{binkert2011gem5}. 
The simulation throughput of a detailed execution-driven simulation is often \san{around five orders of magnitude slower than a real processor.} 
Various statistical approaches have been proposed to accelerate the microarchitecture simulation by only simulating representative~\cite{sherwood2002automatically} or a fraction of instructions~\cite{wunderlich2003smarts} from the benchmark. 
%


\textbf{Motivation (ii)}. Trace-driven simulation faces accuracy concerns in pursuit of higher throughput than execution-driven simulations~\cite{goldschmidt1993accuracy, elrabaa2017very}. 
Figure~\ref{fig:introduction}(b) illustrates this method. For program 1, trace-driven simulation derives the detailed trace of the program on a particular microarchitecture. Subsequently, it simulates that trace on a different microarchitecture and derives the microarchitecture statistics. 
Shade~\cite{cmelik1994shade}, MacSim~\cite{kim2012macsim} and MASE~\cite{larson2001mase} are some of the popular trace-driven simulators. Trace-driven simulation is mostly used to explore the design of specific microarchitecture components like cache and memory. The reference trace captures the events related to CPU core, and cache and memory on $\mu$Arch A. During simulation on a different $\mu$Arch B, the events related to CPU cores are rapidly replayed, while the events related to memory and cache are simulated in detail for $\mu$Arch B.
Replaying the events related to the CPU core is significantly faster than simulating them. Further, because simulating the CPU core-related instructions dominates the overall simulation cost~\cite{butko2015trace}, trace-driven simulation provides decent speedup over the execution-driven counterpart. 
However, reusing the same trace for different microarchitectures leads to accuracy concerns as the execution order of different memory instructions can vary (see~\cite{lee2009accurately}). 


%
%
\textbf{Motivation (iii)}. Emerging DL-based simulations are remarkably fast and can provide comparable cycle-level accuracy, whereas hits three roadblocks: limited output metrics, expensive microarchitecture specific trace generation, and restricted microarchitecture support. Notably, for the DL model of trained microarchitecture, the state-of-the-art DL simulator, i.e., SimNet~\cite{simnet,pandey2022scalable} can achieve >1,000$\times$ higher throughput than the execution-driven simulator, i.e., gem5. 
The substantial increase in throughput is attributed to replacing the highly irregular and heterogeneous simulation with DL models that are accelerator-friendly and parallelizable. 
DL-based simulators, illustrated in Figure~\ref{fig:introduction}(c), typically follow a two-step process to model the performance of a program: Step (i), a DL model specific to a $\mu$Arch A is trained using a detailed trace of a program 1 generated through simulation on $\mu$Arch A. Step (ii), this trained model is generalized to predict output metrics for the traces of unseen programs, i.e., program 2 for the same $\mu$Arch A. However, existing DL-based simulators only predict one output metric, i.e., CPI. Further, because they need low-level microarchitectural performance metrics like branch misprediction and cache misses as the input, one needs to generate microarchitecture specific traces for simulating the same program on different architectures. 
%
When the overhead for generating the trace and training the model is accounted \update{for}, DL-based simulators can be upto 17$\times$ slower than execution-driven simulator gem5 when simulating 1 billion instructions. 


In summary, the mainstream cycle-level microarchitecture simulations excel and fall short at different aspects, i.e., speed, accuracy, and/or simulation details (i.e., \# of output metrics). Particularly, (i) although execution-driven simulation offers the desired accuracy and detail, its slow throughput limits its application scenarios. (ii) Trace-driven simulation achieves decent acceleration by reusing traces but sacrifices accuracy and fails to provide significant speedups. (iii) DL-based simulations promise impressive throughput but fail to provide crucial microarchitectural performance metrics and introduce substantial overheads from trace regeneration and model re-training. 


\subsection{Contributions}
%
%
Departing from the designs and desired goals from the aforementioned three paradigms, this paper redesigns DL-based cycle-level microarchitecture simulator. Particularly, we take as input the functional and detailed traces 
train a DL-based simulator, support \san{a set of desired performance metrics} of interest and fast microarchitecture exploration, achieving the comparable accuracy as execution-driven simulation, and an order of magnitude higher throughput than the state-of-the-art DL-driven simulator, i.e., SimNet. Figure~\ref{fig:introduction}(d) illustrates the workflow of our system, which encompasses the following three contributions: 
 \vspace{.1in}
 
\begin{itemize}

    
    \item First, we introduce a unique training dataset design so that the subsequent simulation (i.e., inference) only needs light-weighted and reusable functional trace as inputs.  
    \vspace{.1in}
    
    \item {Second, for predicting a variety of performance metrics of interest, we propose an DL model with separate embedding and {self-attention based} performance prediction layers.} 
    \vspace{.1in}

    
    \item {Third, we introduce transfer learning techniques to rapidly explore various microarchitectures. This includes architecture agnostic embedding layers and judicious training dataset selection.}
    
\end{itemize}

\section{Background}
\label{sec:background}



\subsection{Architecture simulation}




\textbf{Architecture simulations by level of details.}
Architecture simulators can be divided into functional and detailed simulations based on the level of detail.  
(i) \textit{Functional simulation.}  Functional simulators are designed to model the functionality of a microarchitecture rather than its detailed implementation. They primarily validate hardware functions and generate execution traces for specific workloads. They do not simulate the microarchitecture in detail, so they typically do not produce timing information. However, their lack of detail allows them to operate at a speed that is one to two orders of magnitude faster than a detailed simulation.  
(ii) \textit{Detailed simulation.} Detailed simulators simulate the processor by performing all the operations from each component with cycle granularity. They model the detailed knowledge of how the processor works to capture the dynamic behavior of the microarchitecture, which impacts the performance. It provides a meticulous analysis of performance characteristics, enabling researchers to explore how different microarchitectural elements interact and impact overall performance. While detailed simulation offers higher accuracy, it comes at the cost of increased computational overhead, making it more time-consuming.

\textbf{Execution trace.} This paper extensively uses execution trace, which refers to the stream of instructions generated by functional or detailed simulation. The gem5 simulator is modified to generate execution traces capturing various static instruction properties and dynamic performance metrics. Functional trace refers to the microarchitecture agnostic trace generated with functional simulation using \textit{AtomicSimpleCPU} model. We use the terms functional trace and microarchitecture agnostic trace interchangeably. It only contains static properties like opcode, registers, and other instruction flags. Detailed trace refers to the trace generated with the \textit{O3CPU} model. It captures various microarchitecture specific performance metrics like data access misses, instruction cache misses, branch mispredictions, speculative instructions and latency of individual instructions.

\subsection{Deep learning (DL)-based microarchitecture simulation}
In DL-based approaches, the simulated processor is abstracted as a whole, eliminating the need to simulate individual components within the processor. As DL excels at deriving the sophisticated rules that govern various complex functions, recent work shows that it can capture the microarchitecture simulations in a similar way. 
State-of-the-art DL-based simulations, e.g., SimNet~\cite{simnet} and Ithemal~\cite{mendis2019ithemal} manage to model the performance of a program at the instruction level, generally, in two steps: 
(i) An DL model is trained to capture the complex and dynamic relationships between instructions and the hardware based on the instruction properties and the performance metrics. The performance prediction problem can be defined as below: 
\begin{equation}
    Y = f(x_{0}, x_{1}, ... , x_{n} ),
\end{equation}
where $Y$ is the desired output performance metric of the instruction, typically including the cycles required to execute the instruction. $x_{0}$, $x_{1}$ .. ,$x_{n}$ represent the input features used by the models. 

The input features include the properties of the current and earlier instructions (i.e., context instruction). Context instructions are used to model the dependencies and resource contentions among the instructions. The instruction features of existing DL-based simulators include static features like opcode, registers used by the instruction, branch predictions and data access level. The performance metric and input features are gathered from detailed traces by simulating various programs. $f(.)$ is a microarchitecture specific function the model learns. Earlier state-of-the-art works adapt long short-term memory (LSTM) or convolutional neural networks (CNN) for learning $f(.)$. 
(ii) During inference, this trained DL model can be used to predict the performance metrics of various unseen programs for the same microarchitecture at the instruction level. The required instruction input features are collected from dynamic profiling or simulation for a specific microarchitecture. As the processor is abstracted as a whole, any change in the microarchitecture requires re-training of the DL model with microarchitecture specific training datasets. 

\section{Design Principle, Challenge and Overview}
\label{sec:design}

\textbf{Design Principle \#1}. We advocate that (i) the input to the DL model should only capture the instruction execution sequence and (ii) the DL model should govern the hardware features stemming from the following reasons. First, if the input to the DL model only needs to capture the execution sequence, the DL model captures all the microarchitecture features. That is, a trained DL model can be used to predict the crucial low-level microarchitectural performance metrics (i.e., CPI, branch mispredictions, cache misses) of any benchmarks. Second, generating the microarchitecture agnostic instruction execution sequence of a particular benchmark is significantly faster than generating the traces with architectural information {(We refer to as detailed trace in this paper)}. 
Third, if a microarchitecture practitioner would like to change the microarchitecture of a particular hardware, the inputs to the DL model can be reused.

\textbf{Design Principle \#2.} An DL-based microarchitecture simulator should (i) report various performance metrics during the architecture simulation and (ii) support rapid explorations of different architecture configurations. 
First, existing work primarily provides cycles as the only output metrics from the DL prediction model. One needs to rely on the conventional simulation for the rest of the metrics. \san{This limits the application of existing DL-based simulators for microarchitectural bottleneck analysis.} Further, learning from more metrics helps the model learn more complex program and hardware interactions, improving simulation detail and accuracy. Earlier work has shown that performing multi-metric prediction improves the accuracy of the prediction model~\cite{sykora2022granite}.  
Second, conventional efforts require re-training of the model from scratch. Simulating the whole design would incur huge costs just for the training. This discourages the usability of DL-based architecture simulation. \san{We leverage the proposed microarchitecture agnostic embedding layers for transfer learning and fast adaptation across different microarchitecture designs with a relatively small training dataset.}




 \update{ 
\textbf{Challenges.} {\ours} faces three grand challenges: (i) For the training dataset, we need to associate the microarchitecture impacts with each executed instruction in the functional trace. 
(ii) Reporting various performance metrics demands us to derive sufficiently powerful DL models that can capture the impacts of various hardware components. 
(iii) Training microarchitecture-agnostic program embeddings presents difficulties because the embeddings are biased towards the architecture they are trained on. These three challenges motivate the design of {\ours}. 
 }

\textbf{Overview.} Section~\ref{sec:functional} unveils {\ours}, our multi-modal DL architecture for microarchitecture simulation.
Our approach adheres to design principle \#1 by proposing a workflow to construct training datasets from detailed and functional traces which attributes the differences in these two traces to performance metrics, allowing the reuse of functional traces for varying microarchitectures.
For design principle \#2, we propose multi-metric predictions with feature engineering with a self-attention model to increase the simulation detail. \san{Further, we propose techniques to train microarchitecture agnostic embedding layers that enable fast transfer learning which significantly reduces the re-training overhead of DL-based microarchitecture training and simulation.}





\section{{\ours:} A Fast and Detailed DL {Microarchitecture} Simulator}
\label{sec:functional}

This section discusses three components of {\ours}. First, we present our workflow of training dataset generation. Since we perform supervised learning, we associate microarchitecture agnostic input with microarchitecture specific performance metrics. Our workflow derives this training dataset. 
Second, we introduce our multi-metric ML architecture that takes as input the microarchitecture agnostic inputs and outputs various user-requested performance metrics. Third, we present our microarchitecture agnostic embedding construction for fast transfer learning.


\subsection{Training Dataset Construction}
\label{sec:functional:training}

{\ours} uses functional trace as input to the model 
and the output (i.e. label) {can be various performance metrics. This permits the subsequent simulation (i.e., inference) to only require functional trace as inputs, which can be rapidly generated and reused across microarchitectures. 
For the output, the metrics are instruction latency, branch misprediction, data cache misses, instruction cache misses, and translation lookaside buffer (TLB) misses of each instruction~\cite{zheng2016accurate,lee2006accurate,ardalani2015cross}}. For brevity, we use three major performance metrics, i.e., latency, branch misprediction, and data cache misses, to explain how we process the detail and function traces to arrive at the training dataset. However, it is important to note that {\ours} can \san{potentially} support other performance metrics. Eventually, we introduce an automatic workflow to generate the training dataset for any benchmarks.

\begin{wraptable}[9]{r}{6.5cm}
\small
\centering
\begin{tabular}{|c|cc|}
\hline
             & \multicolumn{2}{c|}{\textbf{\# Detailed vs Functional Trace }} \\ \hline
 &
  \multicolumn{1}{c|}{\begin{tabular}[c]{@{}c@{}}Detailed trace\\ (O3CPU)\end{tabular}} &
  \begin{tabular}[c]{@{}c@{}}Functional trace\\ (AtomicSimpleCPU)\end{tabular} \\ \hline
\textbf{1M}  & \multicolumn{1}{c|}{2,655,925}        & 2,528,617        \\ \hline
\textbf{10M} & \multicolumn{1}{c|}{26,689,939}       & 25,469,667       \\ \hline
\end{tabular}
\vspace{0.1in}
\caption{\# of instructions differences in detailed vs functional trace for 531.deepsjeng\_r benchmark.}
\label{table:count_difference}
\vspace{-0.2in}
\end{wraptable}

Functional and detailed traces output similar sequence order, which permits us to associate each instruction of a functional trace with a detailed trace. However, the challenge is that the difference in number of instructions between detailed and functional traces is nontrivial. 
Table~\ref{table:count_difference} shows the difference in instruction counts for detailed and functional simulations of 531.deepsjeng\_r SPEC 2017 benchmark~\cite{bucek2018spec} for a base ARM microarchitecture. As the table shows, for simulation with 1M and 10M as specified instruction count with gem5, the instruction counts of functional and detailed trace differ in 5.2\% and 4.8\%, respectively.



Detailed trace generally differs from functional trace in the following two aspects. 
First, a detailed trace includes various performance metrics introduced earlier for individual instructions. Second, a detailed trace includes two types of additional dynamic instructions during execution that are missing in the functional trace. Specifically, the detailed trace contains incorrect speculative and stall instructions. Incorrect speculative instructions are the wrongly executed instructions squashed based on branch prediction. Stall instructions are used to stall the pipeline by inserting a no-operation (\texttt{nop}) instruction in the pipeline when any other instructions cannot be executed. 

Our key idea is that both types of additional instructions can be converted into numerical performance differences and attributed to specific instructions from the functional trace. 
Using the stall instructions from the detailed trace as an example, one can project the timing impact of these instructions to the latency of the subsequent instructions. Below, we discuss how we model the impact of speculative and pipeline stall instructions.

\textbf{Squashed speculative instructions.} 
Instructions are speculatively executed following the prediction of whether a conditional branch instruction will be taken or not. If the predicted branch path is correct, speculatively executed {instructions} will be correct, thus the instruction streams of detailed and functional traces will be identical. When a speculative path is wrong due to branch misprediction, speculatively executed instructions should be squashed. This case leads to a distinction between functional and detailed traces. 
Having squashed speculative instructions in the detailed trace avoids separately modeling the total impact of a branch misprediction.  

The total impact of branch misprediction can be accounted for in the functional trace with the fetch timing information obtained from the detailed trace. If a branch is mispredicted, it will delay the fetch of the next correct instruction. In a detailed trace, the fetch latency of the correct instruction does not include the speculation or branch resolution overhead. To include the miss prediction overhead, we remove the squashed instruction from the detailed trace, get the difference in the fetch clock as the fetch latency, and add it to the subsequent instruction. 
 

\textbf{Pipeline stalls.} Stall instructions can be handled similarly to squashed speculative instructions. 
When no instruction can be executed in the pipeline due to dependency or resource contention, \texttt{nop} instructions are filled. Similar to squashed speculative instructions, we remove and project the latency impact of \texttt{nop} instructions to the subsequent instruction. 
We use the fetch clock from the detailed trace to determine the additional fetch latency delay.

  \begin{figure}[h]
	\centering
	\includegraphics[width=\linewidth]{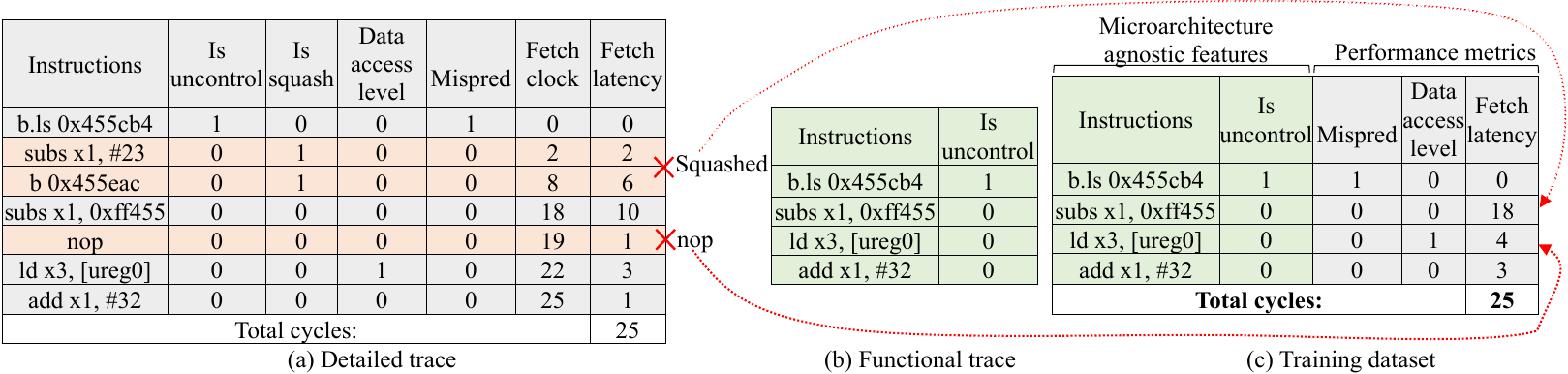}	
    \caption{Training dataset construction illustrated via the trace snippets for 531.deepsjeng\_r benchmark.}   
    \label{fig:trace-difference}
\end{figure}

Figure~\ref{fig:trace-difference} exemplifies that the training dataset resembles functional trace with only modifications regarding performance metrics (see red dashed line arrows). In the detailed trace, the first branch instruction (\texttt{b.ls 0x455cb4}) is mispredicted, and two consecutive instructions are speculatively executed until the branch is resolved. The fetch latency of the next correct instruction (\texttt{subs x1, 0xff455}) is 10, whereas the total overhead of branch misprediction is 18 cycles. 
To model the total impact in training dataset, we remove the speculative instructions and assign the fetch latency of 18 cycles instead of 10 to (\texttt{subs x1, 0xff455}). The fetch latency for (\texttt{subs x1, 0xff455}) changes from 10 to 18 cycles. With the new fetch latency, the ML model can be trained to predict the total impact of misprediction without squashed speculative instructions. Similarly, for stall instructions, the \texttt{nop} instruction is removed from the detailed trace, and then the fetch clock is used to derive fetch latency for the following instruction. The fetch latency for {(\texttt{ld x3, [ureg0]})} is updated to 4 from 3. 
With our workflow, the total cycles remain the same for the detailed trace and adjusted trace, i.e., 25 cycles. \san{In evaluation, we study the differences between detailed and function traces, mainly focusing on speculative and \texttt{nop} instructions for various benchmarks and microarchitectures. }

\subsection{Multi-Metric DL Model Design}
\label{sec:functional:features}
In this section, we delve into feature engineering from the microarchitecture agnostic execution trace and the DL model architecture that predicts various crucial low-level performance metrics using the input features as demonstrated with instruction latencies, branch mispredictions and data access level prediction.  




\begin{figure*}[h]
	\centering
	\includegraphics[width=0.9\linewidth]{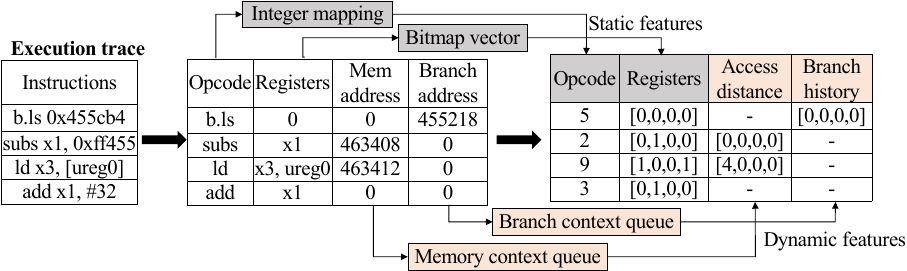}	
    \caption{Feature engineering}
    \label{fig:feature_engineering}
\end{figure*}


\textbf{Feature engineering.} \san{We propose new techniques to build cross-instruction features, in addition to the per-instruction features from the state-of-the-art~\cite{simnet}.}
Figure~\ref{fig:feature_engineering} shows the process of gathering the input features for the model. The input should be representative enough that the DL model can learn to map the interplay between the instruction features and the microarchitecture to predict various performance metrics. We extract four key instruction properties from the microarchitecture agnostic execution trace: the opcode, registers, data access address and PC address. Opcode and registers derive the per-instruction features. For opcode, we employ an integer mapping for each unique opcode in the dataset. Regarding registers, since the instructions can involve multiple registers, we create a bitmap vector with a size equal to the total number of registers. If an instruction uses $i_{th}$ register, $i_{th}$ index in the vector will be set to 1 (0 otherwise).  \update{Both source and destination registers are included in the bitmap vector.}

\begin{wrapfigure}[12]{r}{8cm}
    \vspace{-0.1in}
	\centering
	\includegraphics[width=\linewidth]{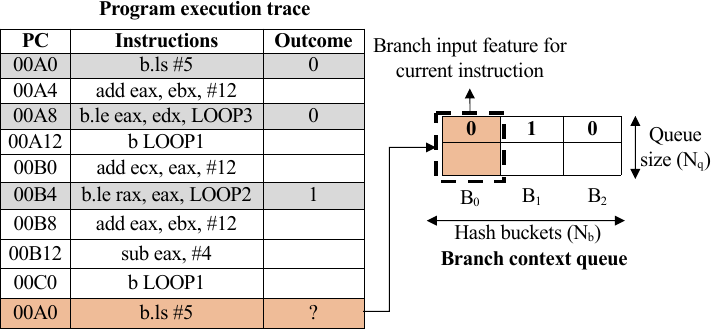}	
    \caption{Input for branch instruction.}
    \label{fig:predictor}
\end{wrapfigure}

Cross-instruction features, crucial for predicting branch misprediction and data access level, are derived from the PC and memory addresses.
We use the branch history as input to model the outcome of conditional branch instructions. This history, indicating the outcomes of prior branch instructions, is employed by existing branch predictors to predict whether the branch will be taken~\cite{hennessy2011computer}. \san{For a given input feature size, storing the outcome of each branch in a separate queue will limit the number of unique branches. To address this, we employ a hash table to store the outcomes of branch instructions. Hashing effectively controls the input feature size while maintaining relevant outcomes history for each branch (exemplified in Figure~\ref{fig:predictor}).}
%


Figure~\ref{fig:predictor} shows an example of retrieving the branch input feature with a sample program execution trace. We construct a hash table with $N_{b}$=3 buckets and $N_{q}$=2. The table is populated as we go through each instruction. When a conditional branch instruction is encountered, we retrieve (PC address\%4$N_{b}$) bucket and use that as branch input features. Subsequently, we update the outcome of that branch to the respective bucket before the next instruction. 
To retrieve the branch input features for the last instruction of the program execution trace, i.e., (\texttt{00A0: b.ls \#5}), we first determine the hash bucket, i.e., (00A0\%4$N_{b}$)=B$_0$. The retrieved branch input feature will be [0 , -]. The input contains the earlier outcome of the same instruction. The hash table effectively separates the outcome of other PC addresses like (\texttt{00A8: b.le eax, edx, LOOP3}) and (\texttt{00B4: b.le rax, eax LOOP2}), which may be unrelated. It is also important to note that this design also permits different branches that are hashed to the same bucket to together offer a global history for future predictions. 



To model the data access level, we calculate the access distance, which is the difference between current memory access and the previous $N_{m}$ memory accesses, and use that as the input to the model. Data access level is used to derive the cache misses. Intuitively, if the access distance between the current and earlier memory access is smaller, current access is more likely to be in the cache. Access distance is similar to reuse~\cite{duong2012improving,kandemir2015memory,zhong2009program,ding2003predicting} or stack distance~\cite{cabetacaval2003estimating,almasi2002calculating,lim2022efficient} histograms in earlier analytical models but cheaper to calculate. We use a memory context queue to track the access distance of $N_{m}$ memory accesses. Figure~\ref{fig:feature_engineering} illustrates how access distance is calculated for memory instructions. In the case of (\texttt{subs x1, 0xff455}), being the first memory access, the access distance is zero. The address is added to the memory context queue. For the second memory access, the difference in memory address with the first instruction is 463412-463408=4. With $N_{m}$=4, the access distance will be [4,0,0,0]. The optimal value of $N_{b}$,$N_{q}$ and $N_{m}$ are empirically derived \update{based on the simulation error across test benchmarks (Section~\ref{sec:eval:dse})}.

\begin{figure*}[h]
	\centering
 \includegraphics[width=\linewidth]{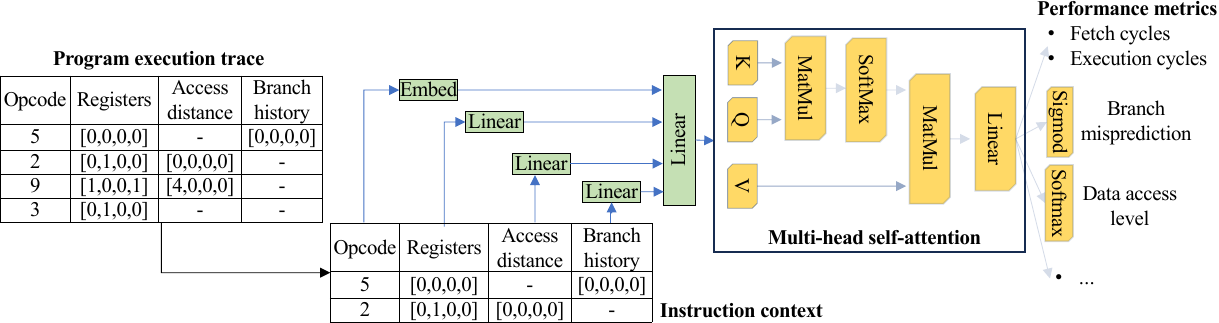}	
    \vspace{-0.1in}
    \caption{Our initial DL model architecture.}   
    \label{fig:overview}
\end{figure*}

\textbf{DL model architecture.} Figure~\ref{fig:overview} exemplifies our DL model design. The model first generates instruction embeddings from input features with two-level embedding layers and then uses multi-headed self-attention to perform multi-metric prediction. We use a sequence of $N$+1 instructions as input to the model. Here, $N$ signifies that earlier instructions can influence the performance of the current instruction, which are the context instructions. Unlike the prior approaches~\cite{simnet,pandey2022scalable} that manages a context instruction queue, adding or removing context instruction based on fetch cycles, our approach relies on the self-attention layer to autonomously learn which earlier instructions significantly impact the current instruction.

The embedding layers generate instruction embeddings in two steps. Initially, embeddings are created independently for each category of input. This separate generation facilitates enhanced representation learning for each category. Specifically, for opcode, a trainable lookup table based embedding layer is employed.
For the remaining categories, distinct linear embedding layers are utilized. The individual instruction embedding is obtained by combining categorical embeddings through a linear layer. Note embedding layers independently generate instruction embeddings for current and $N$ context instructions. Similar to SimNet, we assign the value of $N$ as the maximum value of reorder buffer (ROB) in a design space, \update{in this case,} 128.  

Following the generation of instruction embeddings, the prediction layers employ multi-head self-attention to determine the performance metrics. Considering the impact of microarchitecture, this approach allows attention layers to model the interaction between current and earlier instructions. Using self-attention obviates the need for manually tracking context instructions, enhancing efficiency. Employing multiple heads enables each head to learn unique hardware-instruction interplay. The output from each head is concatenated and passed through a linear layer.


We use different operators to predict different performance metrics based on the output of the last linear layer: (i) 
The fetch and execution cycles are directly predicted from the linear layer. (ii) An additional sigmoid layer is incorporated for branch prediction to predict whether the branch will be mispredicted. (iii) We use a softmax layer for the data access level, as the output can be multiple categories. (iv) More performance metrics like instruction cache miss and TLB miss can be predicted through a sigmoid layer. During training, a loss is computed from each performance metric and combined with a linear ratio in backpropagation. \update{To obtain the total cycle of all instructions, we use the retire clock of instructions. Retire clock is computed as current clock + fetch latency + execution latency. The retire clock of the last instruction of a benchmark determines the total cycles. }

{\bf Intuitive explanation on supporting a set of performance metrics.} Multi-metric prediction exploits the relatedness of performance metrics. With the attention model and microarchitecture agnostic input, our design allows us to output various performance metrics of interest. It can capture the relationship between each performance metric and the specific input features that impact the metric. This allows all metrics to be derived from the same hidden layers. {We demonstrate the validity of this idea by accurately predicting three performance metrics in Section~\ref{sec:eval:ablation}.} Multi-metric prediction has two benefits. First, it increases the output details of the simulation. Second, individual loss from data access level and branch prediction helps the model relate the cycle prediction with memory and branch behavior during training.

\subsection{Fast Transfer Learning via Microarchitecture Agnostic Embeddings} 
\label{sec:functional:embeddings}

\begin{wrapfigure}[13]{r}{5.4cm}
    \vspace{-0.1in}
	\centering
	\includegraphics[width=\linewidth]{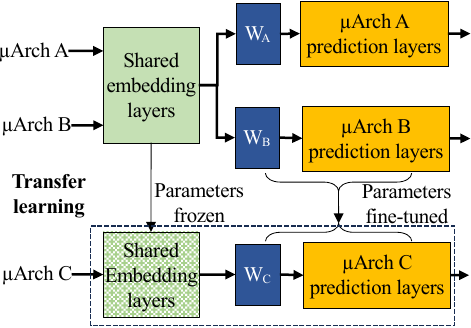}	
    \caption{Overview of \update{transfer learning process for microarchitecture $\mu$Arch C}.}
    \label{fig:TL}
\end{wrapfigure}

Figure~\ref{fig:TL} illustrates our fast transfer learning process to enable {\ours} for a new \update{unseen} microarchitecture rapidly, i.e., $\mu$Arch C, employing microarchitecture agnostic embedding layers and fine-tuning. Initially, shared embedding layers are trained with two carefully selected microarchitectures, i.e., $\mu$Arch A and $\mu$Arch B. 
During training for $\mu$Arch C, the parameters of shared embedding layers are frozen, i.e., we do not update the parameters during backpropagation. The parameters of prediction layers and embedding adaptation layer are fine-tuned with the training dataset for $\mu$Arch C.

\textbf{Microarchitecture agnostic embedding design.} 
The shared embedding layers generate embedding for each individual instruction, and microarchitecture specific prediction layers predict the performance labels. The prediction layers of each microarchitecture computes the gradients for the embedding layers separately. We propose to combine them to update the shared embedding layers. 

Such designs that combine gradients to update shared layers can face two critical issues: \textit{negative transfer} and \textit{imbalance in gradient magnitude} for shared layers: (i) {Negative transfer~\cite{zhao2018modulation,liu2019loss} occurs when the shared layers receive gradients from different microarchitecture that are opposite to each other}. (ii) Imbalance in gradients magnitude~\cite{chen2018gradnorm} arises when one microarchitecture is too dominant during training, inducing gradients with relatively large magnitudes. These issues impact convergence and generalization~\cite{zhao2018modulation,chen2018gradnorm}.

\begin{figure}[h]
	\centering
        \includegraphics[width=\linewidth]{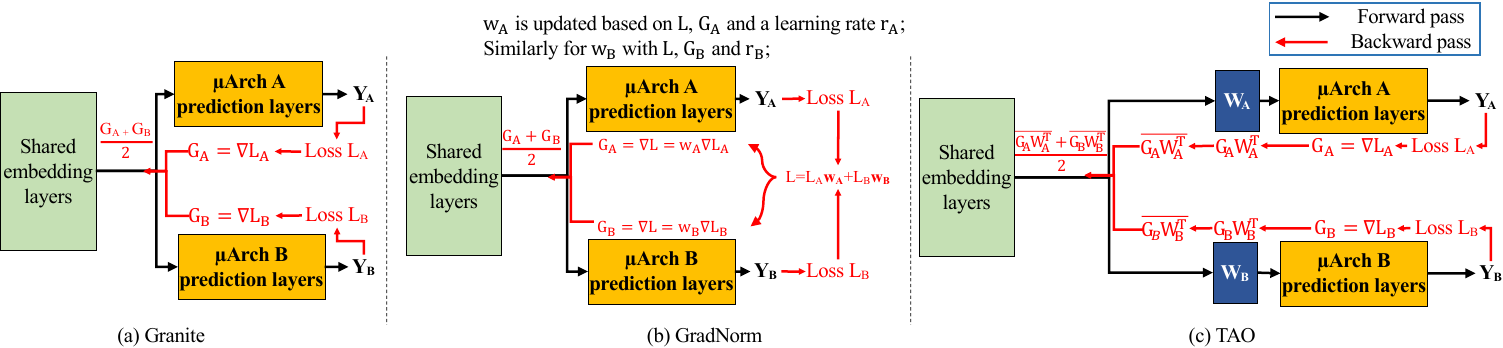}	
    \caption{Comparison of multi-architecture training paradigm.}
    \label{fig:tl_design}
\end{figure}

Figure~\ref{fig:tl_design} compares our multi-architecture training paradigm with two existing projects, Granite~\cite{sykora2022granite} and GradNorm~\cite{chen2018gradnorm}. We use two microarchitectures A and B, to illustrate the techniques. Although GradNorm is proposed for multi-task learning, we compare its effectiveness for generating microarchitecture agnostic embedding layers. In the figure, each prediction network predicts microarchitecture specific output labels ($Y_{A}$ and $Y_{B}$) and losses ($L_{A}$ and $L_{B}$).

In Granite, Figure~\ref{fig:tl_design}(a), to derive the gradients for shared embedding layers, the gradients from the prediction layers of each $\mu Arch$ are averaged (i.e., $\frac{G_{A} + G_{B}}{2}$). Just averaging the gradients may resolve neither the negative transfer nor gradient imbalance problem~\cite{yu2020gradient}. Using gradient imbalance as an example, if the gradient of one task is larger in magnitude than the other, the larger one will dominate the average gradients.

GradNorm, Figure~\ref{fig:tl_design}(b), addresses the imbalance in gradient magnitude for multi-task learning by using learnable combination weights ($w_{A}$ and $w_{B}$) to combine the losses from each task. This indirectly controls the magnitude of the gradients. The underlying rationale is to dynamically adjust the combination weights in response to the gradient magnitudes of shared layers, ensuring they neither become excessively large nor too small. The process begins with the computation of a combined loss ($L$) as a weighted sum of microarchitecture specific weights and loss, i.e., $L_{A}w_{A} + L_{B}w_{B}$. Subsequently, a standard backward pass generates gradients $G_{A}$ and $G_{B}$ for the respective prediction layers using $L$. $G_{A}$ and $G_{B}$ are basically weighted gradient loss i.e., $w_{A}\nabla L_{A}$ and $w_{B}\nabla L_{B}$, respectively. 
$G_{A}$ and $G_{B}$ are averaged for computing gradients of embedding layers. Combination weights are updated based on L, $G_{A}$ and a learning rate $r_{A}$, described in~\cite{chen2018gradnorm}. In this way, GradNorm indirectly balances the magnitude of gradients by updating the loss weight, i.e., $w_{A}$ and $w_{B}$ for various tasks. 

While GradNorm can effectively address gradient magnitude imbalance, it cannot adequately address negative transfer issues that arise from conflicting gradient directions. Of note, conflicting gradients may appear when the performances of two different microarchitectures are opposite for the same instruction. Modifying the magnitude of gradients may not effectively change gradient direction in joint training~\cite{yu2020gradient}. Hence, it may not fully mitigate the adversarial effect of gradients.

Figure~\ref{fig:tl_design}(c) illustrates our design that tackles negative transfer and gradient imbalance. In contrast to GradNorm which relies on reactive approaches of projecting conflicting gradients to a different plane~\cite{yu2020gradient} or finding common direction~\cite{sener2018multi} to mitigate negative transfer, we adopt a proactive solution. We add an individual embedding adaptation layer, i.e., $W_{A}$ for $\mu$Arch A, similarly $W_{B}$ for $\mu$Arch B, between the embedding and performance network, see Figure~\ref{fig:tl_design}(c). The linear layer $W_{A}$ projects the shared embedding (i.e., Green layers) into microarchitecture specific spaces (i.e., $\mu$Arch prediction layers) during forward propagation.

Adding this linear projection layer resolves the negative transfer issue as follows: during backpropagation, to compute the gradients for the linear projection layer, we multiply the gradients from the earlier layer $G_{A}$ with the transpose of the weight matrix $W_{A}$, i.e.,  $G_{A}W^{T}_{A}$ based on the chain rule. 
Under most of the cases, this operation rotates the gradients in the gradient space, changing the direction of gradients. 
For a very rare case when all the columns of $G_{A}$ are the eigenvectors of $W^{T}_{A}$, this linear projection does not change the direction. This requires all the columns of both $G_{A}$ and $G_{B}$ to, respectively, be the eigenvectors of $W^{T}_{A}$ and $W^{T}_{B}$ to render our method in vein. We regard this scenario to be extremely rare. In this paper, for 200 epochs across three different microarchitectures also suggests that we do not experience such a rare case.


To tackle the gradient imbalance concern, we normalize the gradients for the embedding layers based on the magnitude of the gradients $\overline{G_{A}W_{A}}$ and $\overline{G_{B}W_{B}}$ to reduce any existing gradient magnitude imbalance. 
We adopt a typical normalization method: we first compute the mean $X_{mean}$ of a gradient matrix $X$. Then, we get the difference of the gradient matrix with its mean (X-${X_{mean}}$). The difference is divided by the range of the values in the gradient matrix, i.e.,  $\frac{X-X_{mean}}{X_{max}-X_{min}}$. We perform this normalization individually for each gradient matrix. This normalization ensures that both gradients have the similar scale.
The average of normalized gradients, i.e., $ \frac{\overline{G_{A}W^{T}_{A}} + {\overline{G_{B}W^{T}_{B}}}} {2}$ is used to update the shared embedding layers.

\begin{algorithm}[t]
\begin{algorithmic}[1]
\Statex  Initialize model weights, along with $W_{A}$ and $W_{B}$ \hfill
\For{each epoch} 
\State Compute $L_{A}$ and $L_{B}$  \Comment{Standard forward pass}    
\State Compute gradients $G_{A}$ and $G_{B}$ for the prediction layers     
\State Compute gradients $G_{A}W^{T}_{A}$ and $G_{B}W^{T}_{B}$ for the linear projection layer  
\State Normalize the gradients: $\overline{G_{A}W^{T}_{A}}$ $\gets$ normalize($G_{A}W^{T}_{A}$), and $\overline{G_{B}W^{T}_{B}}$ $\gets$ normalize($G_{B}W^{T}_{B}$)  
\State Compute average of normalized gradients $\frac{\overline{G_{A}W^{T}_{A}} + {\overline{G_{B}W^{T}_{B}}}} {2}$      
\State Use the average of normalized gradients to update embedding layers and update model parameters  
\EndFor
\end{algorithmic}
\caption{Training workflow for shared embedding layers with {\ours}}
\label{algorithm:train}
\end{algorithm}

Algorithm~\ref{algorithm:train} explains the workflow. First, microarchitecture specific loss $L_{A}$ and $L_{B}$ are computed. The gradient for each performance prediction layer $G_{A}$ and $G_{B}$ is calculated based on $L_{A}$ and $L_{B}$, respectively. Then, we calculate the gradients for the linear projection layer as $G_{A}W^{T}_{A}$ and $G_{B}W^{T}_{B}$. For gradient normalization, we normalize both gradients individually, $\overline{G_{A}W^{T}_{A}}$ and $\overline{G_{B}W^{T}_{B}}$. The final gradients for the embedding layers will be the average of normalized gradients, $\frac{\overline{G_{A}W^{T}_{A}} + {\overline{G_{B}W^{T}_{B}}}} {2}$. Finally, we update the gradients for embedding layers and continue the backward pass. 

\textbf{Training dataset}. 
{\ours} only uses two microarchitectures based on performance variations to train the model efficiently with the desired accuracy. This is significantly more efficient than training general embedding layers with random microarchitectures. To achieve the accuracy and efficiency goal, we define metrics to measure the architectural variations and select the two architectural variations with the most difference. Below are our designs:

To measure the microarchitecture variations, we select four performance metrics, i.e., CPI, L1 cache miss, L2 cache miss, and branch misprediction rate. 
Of note, since our embedding is performance embedding, we tie the microarchitecture variation to performance metrics. 
We choose these four performance metrics because they can capture the processor, cache, memory, and branch behaviors of a program. Combinedly, these metrics explain the performance impact of key microarchitecture components on overall performance. The choice is also evident by the earlier project~\cite{fields2003using, sherwood2002automatically, lee2006accurate}, which solely uses these metrics to perform microarchitectural bottleneck analysis and hardware design space exploration. 


We measure the performance metrics difference of different microarchitectures with Mahalanobis distance~\cite{mclachlan1999mahalanobis} instead of Euclidean or Cosine distance for two reasons: (i) Euclidean distance is sensitive to a larger value of one metric, and Cosine distance ignores the value difference. (ii) The other two distances do not consider the correlation among the performance metrics or their scales during distance computation. 
Mahalanobis distance measures the distance between two points in a multi-dimensional space. For two vectors X and Y, Mahalanobis distance is defined as $D_{MD}(X, Y) = \sqrt{(X - Y)^T \cdot S^{-1} \cdot (X - Y)}$, where $S^{-1}$ represents the inverse of the covariance matrix of performance metrics from all designs. The covariance matrix represents how the performance metrics vary together across the designs. Using the inverse of the covariance matrix, Mahalanobis distance normalizes the data and accounts for the correlation between dimensions. This normalization makes it less sensitive to a larger metric value.

\begin{figure}[h]
	\centering
        \includegraphics[width=\linewidth]{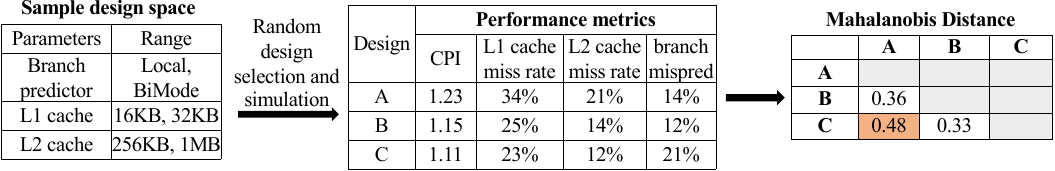}	
    \caption{Selecting training dataset.}
    \label{fig:tl_dataset}
\end{figure}

Figure~\ref{fig:tl_dataset} shows an example of the overall process. First, we randomly select N designs from the design space. 
Here, we select three designs A, B and C. We perform detailed simulations of those designs using gem5 to gather the performance metrics. 
%
%
The performance metric is averaged across the benchmarks. Then, we calculate the Mahalanobis distance for all designs, resulting in a 3x3 matrix. Based on Mahalanobis distance, we select two designs with the largest distance. Here, as the distance between A and C is the largest among all pairs, i.e., 0.48, we select A and C for microarchitecture agnostic embedding construction. 
Of note, since selecting a training dataset is one time cost, the overhead can be considered as preprocessing time.

\section{Evaluation}
\label{sec:eval}

\begin{table}[h]
\small
\begin{tabular}{|l|c|c|}
\hline
 &
  \textbf{Datasets} &
  \textbf{Abbr.} \\ \hline
\textbf{Training} &
  531.deepsjeng\_r, 654.roms\_s,  544.nab\_r, 641.leela\_s &
  dee, rom, nab, lee \\ \hline
\textbf{Testing} &
  605.mcf\_s,  523.xalancbmk\_r, 621.wrf\_s, 507.cactuBSSN\_r &
  mcf, xal, wrf, cac \\ \hline

\end{tabular}
\vspace{0.05in}
\caption{SPEC CPU2017 benchmarks used for training and testing.}
\label{table:datasets}
\vspace{-0.1in}
\end{table}

\textbf{Benchmarks.}
We use the widely adopted SPEC CPU2017~\cite{bucek2018spec} benchmark suite to evaluate \ours. The benchmark suite contains various benchmarks designated for ``speed'' or ``rate'' for INT and FLOAT workloads, resulting in a diverse and complex range of applications like 3D rendering, image manipulation, compression, etc. We select a subset of benchmarks from each category to train and evaluate the model. Instead of randomly selecting train/test benchmarks, we select unique representative benchmarks based on the performance variations as suggested by~\cite{panda2018wait} (see Table V). This allows the model to be trained by diverse instructions from various benchmarks and helps generalize {\ours} over new benchmarks. Table~\ref{table:datasets} shows our training and testing datasets. For a fair comparison, this train/test dataset is used for all related evaluations.

\textbf{Training dataset.} To construct the training dataset, we first generate detailed and functional traces with {100 million instructions} from each training benchmark with default test workloads using the gem5 O3CPU and AtomicSimpleCPU model, respectively. Of note, we skip the first 100 million instructions as adopted by earlier projects to avoid the common program initialization phase~\cite{sherwood2002automatically, dubach2007microarchitectural}. The benchmarks are compiled and simulated in ARM Instruction Set Architecture (ISA). For preprocessing, we remove the duplicate samples from the dataset and generate input features with workflow as discussed in Section~\ref{sec:functional:training}. After preprocessing, the resulting training dataset contains around 180 million instructions across four training benchmarks (See Table~\ref{table:datasets}).

%


\begin{table}[h]
\small
\centering
\setlength{\tabcolsep}{2pt}
\begin{tabular}{|c|c|c|c|c|c|c|c|}
\cline{1-3} \cline{5-7}
\textbf{Components} &
  \textbf{\begin{tabular}[c]{@{}c@{}}Design \\ parameters\end{tabular}} & \textbf{Range} & & \textbf{$\mu$Arch A} & \textbf{$\mu$Arch B} & \textbf{$\mu$Arch C}\\ \cline{1-3} \cline{5-7}
\multirow{2}{*}{Pipeline}  & Fetch width     & 2,3,4                                                                & & 2 & 3 & 4 \\ \cline{2-3} \cline{5-7}
                           & ROB size        & 32, 64, 96, 128                                                        & & 32 & 96 & 128\\ \cline{1-3} \cline{5-7}
{\begin{tabular}[c]{@{}c@{}}Branch pred.\end{tabular}} &
  Algorithm &  Local, BiMode, TAGE\_SC\_L, Tournament  & & Local & BiMode & Tournament  \\   \cline{1-3} \cline{5-7} 
\multirow{2}{*}{L1 Dcache} & Associativity        & 2, 4, 6, 8                                                      & & 2 & 4 & 8\\ \cline{2-3} \cline{5-7}
                           & Size            & 16KB, 32KB, 64KB, 128KB & & 16KB & 32KB & 64KB\\ \cline{1-3} \cline{5-7}
\multirow{2}{*}{L1 Icache} & Associativity       & 2, 4, 6, 8                                                      & & 2 & 4 & 8\\ \cline{2-3} \cline{5-7}
                           & Size            & 8KB, 16KB, 32KB                                                         & & 8KB & 16KB & 32KB \\ \cline{1-3} \cline{5-7}
\multirow{2}{*}{L2 Dcache} & Associativity       & 2, 4, 6, 8                                                      & & 2 & 4 & 8\\ \cline{2-3} \cline{5-7}
                           & Size            & 256KB, 512KB, 1MB, 2MB, 4MB & & 256KB & 1MB & 4MB\\ \cline{1-3} \cline{5-7}
\end{tabular}
\vspace{0.1in}
\caption{Microarchitectural design space parameters choices.}
\label{table:design_space}
\vspace{-0.2in}
\end{table}

\textbf{Design space.}
Table~\ref{table:design_space} shows the overall design space and microarchitecture designs used for evaluations in the paper.
We vary various microarchitecture parameters related to the pipeline, cache and branch predictors, similar to those other researchers have looked for evaluations~\cite{dubach2007microarchitectural, van2016analytical, eyerman2009mechanistic, joseph2006predictive}. We select nine design parameters with varying ranges for a single-core superscalar CPU. For example, the ROB has a minimum size of 32 entries and a maximum size of 128 entries. For evaluating simulation accuracy and throughput, we select three microarchitecture designs ($\mu$Arch A, $\mu$Arch B and $\mu$Arch C) with large variations from Table~\ref{table:design_space} to demonstrate the robustness of our approach. The microarchitecture parameters for each design are also shown in the table. Each microarchitecture is evaluated on the test benchmarks in Table~\ref{table:datasets}. A separate DL model is trained for each microarchitecture design with transfer learning (see Figure~\ref{fig:TL}).

\textbf{Simulation study criteria.} 
We study the simulation error for CPI, branch prediction, and memory access levels, and throughput in this section. Particularly, simulation error represents the absolute CPI prediction error for each benchmark and is defined as $\frac{|CPI_{pred}-CPI_{truth}|}{CPI_{truth}}\times$ 100\% . CPI is calculated by the sum predicted cycle of all instructions divided by the total count of instructions. $CPI_{pred}$ and $CPI_{truth}$ represents the CPI derived from {\ours}  gem5, respectively. For evaluating cache misses and branch misprediction accuracy, we use misses per kilo instructions (MPKI). We use \texttt{\{<microarchitecture name>- <benchmark name>\}} notation in the plots to represent the outcome of benchmark \texttt{<benchmark name>} on microarchitecture \texttt{<microarchitecture name>}. 
Simulation throughput is measured in million instructions per second (MIPS).
For evaluating {\ours}, we generate a functional trace with 100 million instructions for each test benchmark using gem5 AtomicSimpleCPU model, similar to earlier studies~\cite{dubach2007microarchitectural, simnet}.

\textbf{System.}
 For training and simulation, we evaluate our work on a server with four A100 GPUs (80 GB) and an Intel(R) Xeon(R) Silver 4309Y 32-core CPU. \update{We use GPUs for the DL model inference as it provides significantly higher throughput than CPUs.} Other hardware accelerators like FPGA can also be used for model inference~\cite{li2020ftrans}. For comparison with the state-of-the-art work, i.e., SimNet, we use the CNN (C3 hybrid) model as described in the paper~\cite{simnet} and GitHub~\cite{simnetgithub}. We use the same datasets to train SimNet and {\ours}. The models are trained with Pytorch 2.1.0. 


\subsection{Comparison with the State-of-the-Art}
\begin{figure}[h]
    \centering
        \includegraphics[width=\linewidth]{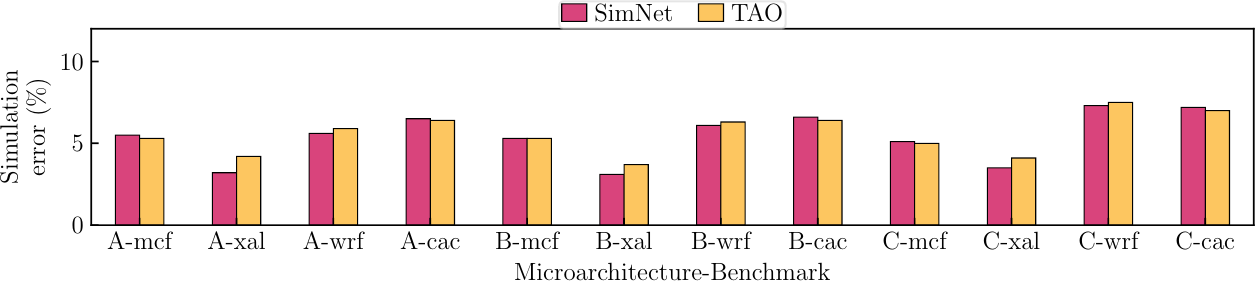}
    \caption{Simulation accuracy comparison with the state-of-the-art.} 
    \label{fig:sota} 
    \vspace{-0.1in}
\end{figure}

This section compares the simulation accuracy and overall simulation time of {\ours} with the state-of-the-art DL-based simulator SimNet.

Figure~\ref{fig:sota} compares the simulation error for the selected three microarchitectures and four test benchmarks. The x-axis represents the simulation error derived from gem5, and the y-axis represents benchmarks from different microarchitectures. In most microarchitectures and benchmarks, {\ours} closely matches the simulation error of SimNet. On average, SimNet and {\ours} exhibit simulation errors of 5.11\% and 5.23\%, respectively. {The slightly higher simulation error of TAo can be attributed to prediction error for branch misprediction and cache misses.}  Interestingly, {\ours} performs relatively better in \texttt{mcf} and \texttt{cac} benchmark. The improvement in \texttt{mcf} can be attributed to relatively higher arithmetic instructions of \texttt{mcf} benchmark, in which {\ours} can provide better prediction with an embedding representation of instructions while SimNet uses numerical representation. \texttt{cac} has a relatively higher number of memory stores and fewer branch instructions. SimNet incurs higher errors for memory store instructions due to limited input features relating to memory store instructions. For the evaluated microarchitectures and benchmarks, {\ours} demonstrates a maximum and minimum simulation error of 3.7\% and 7.4\%, respectively. Benchmark \texttt{cac} has a relatively higher simulation error than other benchmarks for both SimNet and {\ours}, likely due to its distinctive behavior in memory access among all the benchmarks. Notably, {\ours} maintains similar accuracy as SimNet without performance metrics across benchmarks and microarchitecture designs.

\begin{table}[h]
\small
\centering
\begin{tabular}{|cc|c|c|cc|c|c|}
\hline
\multicolumn{2}{|c|}{} &
  \textbf{\ours} &
  \textbf{SimNet} &
  \multicolumn{2}{c|}{\begin{tabular}[c]{@{}c@{}}Speedup\\ (vs. SimNet)\end{tabular}} &
  \textbf{gem5} &
  \begin{tabular}[c]{@{}c@{}}Speedup\\ (vs. gem5)\end{tabular} \\ \hline
\multicolumn{2}{|c|}{Training}     & 1.9 hours    & 54.2 hours     & \multicolumn{2}{c|}{28.52$\times$}    & -         & - \\ \hline
\multicolumn{1}{|c|}{\multirow{2}{*}{Simulation}} &
  Trace generation &
  0.53 hours &
  13.22 hours &
  \multicolumn{1}{c|}{24.94$\times$} &
  \multirow{2}{*}{7.81$\times$} &
  \multirow{2}{*}{14.01 hours} &
  \multirow{2}{*}{7.26$\times$} \\ \cline{2-5}
\multicolumn{1}{|c|}{} & Inference & 1.41 hours & 1.93 hours & \multicolumn{1}{c|}{1.37$\times$} &  &           &   \\ \hline
\multicolumn{2}{|c|}{Overall}      & 3.84 hours   & 69.35 hours  & \multicolumn{2}{c|}{18.06$\times$}   & 14.01 hours & 3.66$\times$ \\ \hline
\end{tabular}
\vspace{0.1in}
\caption{\update{Simulation time comparison with the state-of-the-art DL-based simulator SimNet and traditional simulator gem5 for 10 billion instructions.}} 
\vspace{-0.1in}
\label{table:overall}
\end{table}

Table~\ref{table:overall} compares the overall time for training and simulation of {SimNet} vs {\ours}. Both SimNet and {\ours} are trained until the error during training is under 6\%. It takes 54.2 hours to train a CNN SimNet model. Meanwhile, with microarchitecture agnostic embeddings and transfer learning (Section~\ref{ref:eval:tl}), {\ours} can train a model with similar accuracy in merely 1.9 hours. 
It improves the training time by 28.52$\times$. \update{Overhead associated with transfer learning for {\ours} is discussed in Section~\ref{ref:eval:tl}.}

For simulation, SimNet requires 13.22 hours to generate an input trace with 10 billion instructions. 
In contrast, utilizing the microarchitecture independent trace which do not simulate any microarchitecture component, the trace generation time is significantly reduced to 0.53 hours for {\ours}.
\update{{\ours}, along with SimNet, performs parallel simulation to provide highly scalable simulation throughput. We follow the parallel simulation technique described in~\cite{pandey2022scalable}. The program traces are partitioned into subtraces and simulated in parallel. }
For SimNet, it takes 1.93 hours to simulate 10 billion instructions with a simulation throughput of 1.46 MIPS. On the other hand, {\ours} completes the simulation in 1.41 hours with a throughput of 1.98 MIPS. This speedup results from two aspects: (i) {\ours} only needs to generate a functional trace for simulation, which is 24.94$\times$ faster. Of note, the functional trace is architecture agnostic, which implies we can potentially avoid trace generation from one microarchitecture to another. \update{The input trace of SimNet requires simulation of cache, and branch along with additional simulation of pipeline. Furthermore, for each microarchitecture change, the trace needs to be regenerated.} (ii) During inference, the relatively slow throughput of SimNet is attributed to history context simulation involving frequent CPU-GPU data movements. The simulation throughput for {\ours} can be further improved with various self-attention optimization techniques~\cite{chen2021re,guo2020accelerating} and further simulation optimizations discussed in~\cite{pandey2022scalable}. Leveraging functional trace and efficient DL-based simulation workflow, the simulation process is accelerated by 7.81$\times$. Overall, {\ours} demonstrates a remarkable speed advantage for simulating a new microarchitecture, being 18.06$\times$ faster than SimNet. This speedup is linearly scaled with the number of microarchitecture designs and benchmarks used to simulate. \update{gem5 provides a simulation throughput of 0.198 MIPS and takes 14.01 hours to simulate 10 billion instructions. {\ours} provides 7.26$\times$ speedup for simulation against gem5. Even including the training time, $\ours$ provides a speedup of 3.66$\times$. TAO provides further speedup when simulating more instructions or using more GPUs for simulation.} 


\subsection{Detailed Trace and Functional Trace for Training Dataset}

\begin{figure}[h]
    \vspace{-0.1in}
    \centering

     \subfloat[Instruction differences in percentage for speculative and nop instructions. ]{
        \hspace{-.1in} \includegraphics[width=0.49\linewidth]{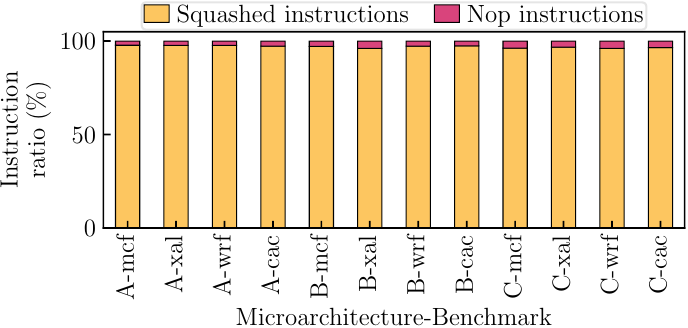}
    } \hspace{.1in}
     \subfloat[Trace generation throughput comparison for detailed vs functional traces. ]{
        \hspace{-.1in}\includegraphics[width=.49\linewidth]{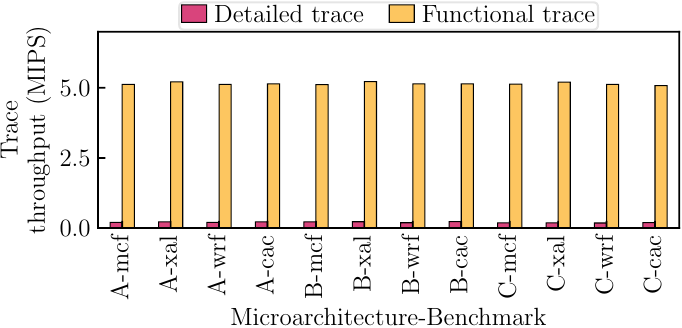}
    } 
    \caption{Choice of the context size and branch configuration.}
    \label{fig:trace}
    \end{figure}

Figure~\ref{fig:trace}(a)  analyzes the ratio of instruction differences in the detailed trace compared to the functional trace used for training dataset construction. The y-axis represents the instruction ratio of each instruction type, and the x-axis represents the microarchitectures and benchmarks. For 100 million simulated instructions from each benchmark, the detailed trace has, on average 96.98\% of squashed pipeline instructions and 3.02\% of nop instructions. The remaining instructions remain the same for different microarchitectures. The variation in the count of speculative instructions across benchmarks comes from different branch predictors and their respective accuracy.

Figure~\ref{fig:trace}(b) compares the trace generation throughput for the detailed trace and the functional trace used by {\ours} for simulation. On average, the trace generation throughput for detailed and functional traces is 0.21 and 5.29 MIPS, respectively. The functional traces utilized by {\ours} exhibit a remarkable speed advantage, being generated 25.19 $\times$ faster than their detailed counterparts. The slower throughput of detailed traces can be attributed to the intricate modeling of various hardware components, such as memory, cache, and branch predictor. Notably, the functional trace throughput remains consistent across different microarchitectures for the same benchmarks. $\mu$Arch A, characterized by a higher occurrence of branch mispredictions, exhibits an elevated number of speculative instructions. Consequently, the average trace throughput for $\mu$Arch A (0.19 MIPS) is marginally lower compared to $\mu$Arch B (0.21 MIPS) and $\mu$Arch C (0.23 MIPS).

\begin{figure}[h]
    \vspace{-0.1in}
    \centering
     \subfloat[mcf:CPI.]{
        \hspace{-.1in} \includegraphics[width=0.32\linewidth]{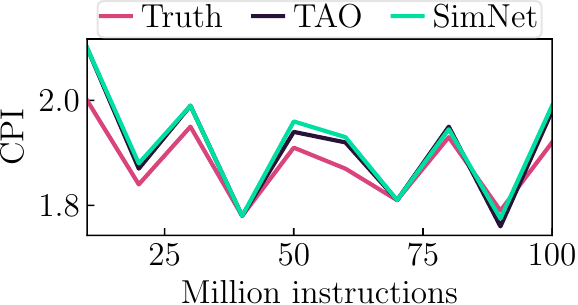}
    } \hspace{.05in}
     \subfloat[mcf:L1 Dcache misses.]{
        \hspace{-.1in} \includegraphics[width=0.32\linewidth]{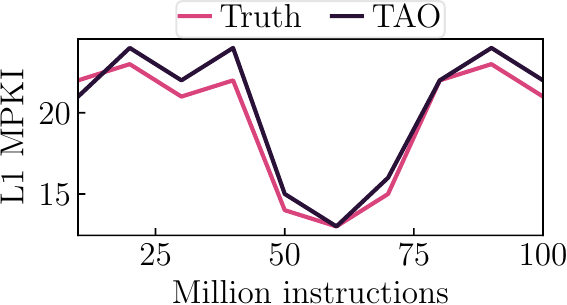}
    } \hspace{.05in}
     \subfloat[mcf:Branch mispredictions.]{
        \hspace{-.1in}\includegraphics[width=.32\linewidth]{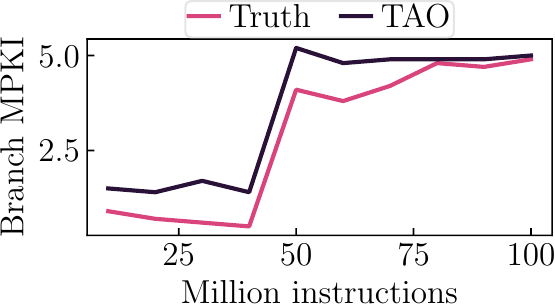}
    }  \\
     \subfloat[xal:CPI.]{
        \hspace{-.1in} \includegraphics[width=0.32\linewidth]{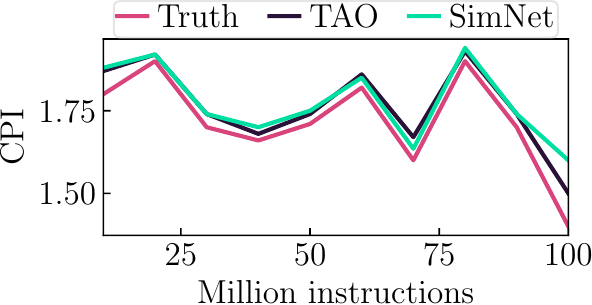}
    } \hspace{.05in}
     \subfloat[xal:L1 Dcache misses.]{
        \hspace{-.1in} \includegraphics[width=0.32\linewidth]{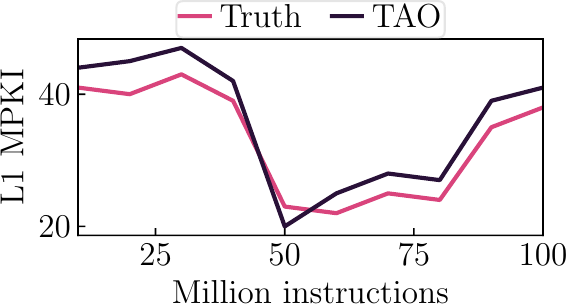}
    } \hspace{.05in}
     \subfloat[xal:branch mispredictions.]{
        \hspace{-.1in}\includegraphics[width=.32\linewidth]{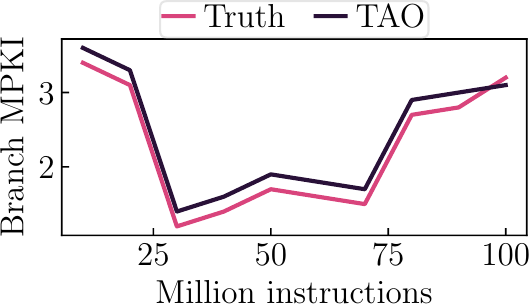}
    }  \\
     \subfloat[wrf:CPI.]{
        \hspace{-.1in} \includegraphics[width=0.32\linewidth]{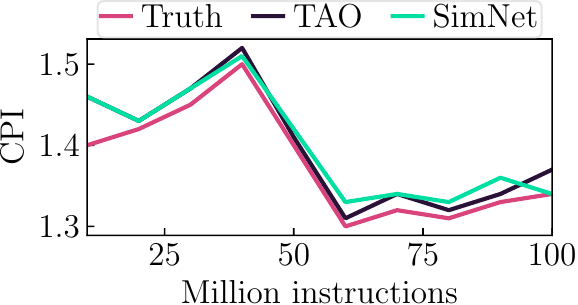}
    } \hspace{.05in}
     \subfloat[wrf:L1 Dcache misses.]{
        \hspace{-.1in} \includegraphics[width=0.32\linewidth]{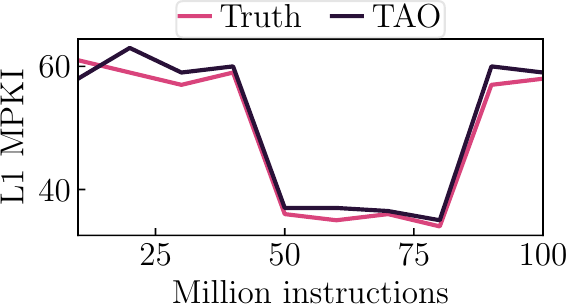}
    } \hspace{.05in}
     \subfloat[wrf:branch mispredictions.]{
        \hspace{-.1in}\includegraphics[width=.32\linewidth]{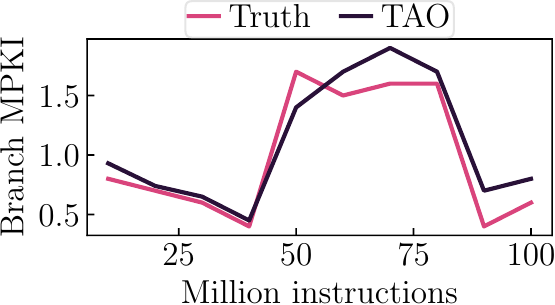}
    }  \\
     \subfloat[cac:CPI.]{
        \hspace{-.1in} \includegraphics[width=0.32\linewidth]{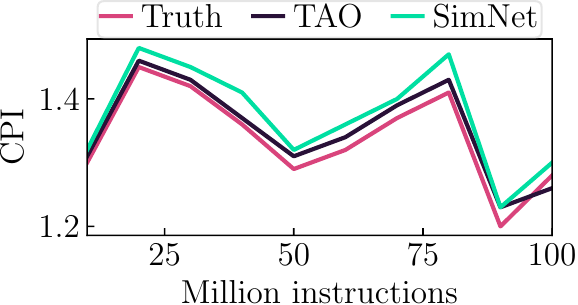}
    } \hspace{.05in}
     \subfloat[cac:L1 Dcache misses.]{
        \hspace{-.1in} \includegraphics[width=0.32\linewidth]{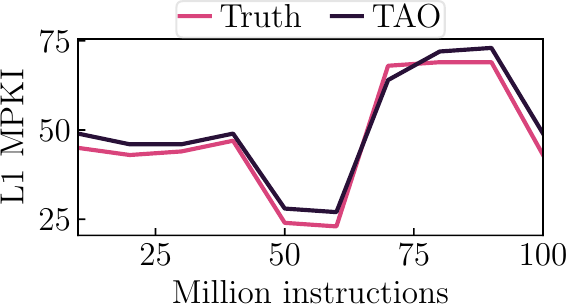}
    } \hspace{.05in}
     \subfloat[cac:branch mispredictions.]{
        \hspace{-.1in}\includegraphics[width=.32\linewidth]{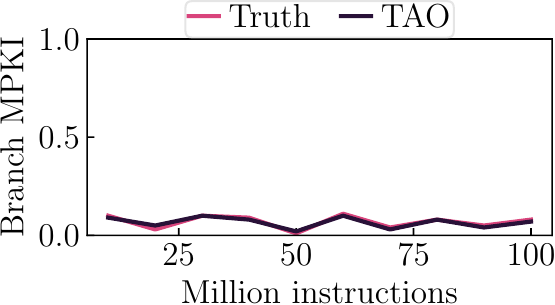}
    }  
    \caption{Phase behavior for test benchmarks.}
    \label{fig:phase}
    \vspace{-0.1in}
    \end{figure}


\subsection{Phase Level Behavior}
Figures~\ref{fig:phase} shows the phase level behavior for benchmarks \texttt{mcf}, \texttt{xal}, \texttt{wrf} and \texttt{cac}, respectively. We compare the CPI, L1 misses and branch misprediction for each benchmark against the ground truth generated from the gem5 simulation for microarchitecture $\mu$Arch A. While SimNet only captures the phase level behavior of CPI, {\ours} can also capture the behavior of instruction cache misses and branch mispredictions. Hence, we only compare against CPI for SimNet. The y-axis for L1 Dcache misses and branch mispredictions in the figure represents MPKI. We compute average CPI, L1 misses, and branch misprediction per ten million instructions. The x-axis shows the number of instructions in millions.

Our evaluation reveals that {\ours} adeptly captures the dynamic behavior of the program for each performance metric during execution. For CPI, {\ours} accurately captures performance variation across different phases of program execution. Notably, {\ours} shows slightly better phase level prediction for \texttt{cac} benchmark (Figure~\ref{fig:phase}(j)), attributed to its enhanced accuracy in predicting latency for store instructions within the benchmark.
For L1 Dcache misses, {\ours} precisely captures the behavior for most of the benchmarks. 
On average, {\ours} shows slightly higher prediction error for branch MPKI than CPI and L1 Dcache MPKI. Nevertheless, it still effectively captures the trend in branch MPKI over the course of program execution. Benchmark \texttt{cac} does not show much variation in branch MPKI due to a lower count of branch instructions (Figure~\ref{fig:phase}(l)).


\begin{figure}[h]
    \vspace{-0.1in}
    \centering

     \subfloat[Varying $N_m$ for data access level input. ]{
        \hspace{-.1in} \includegraphics[width=0.49\linewidth]{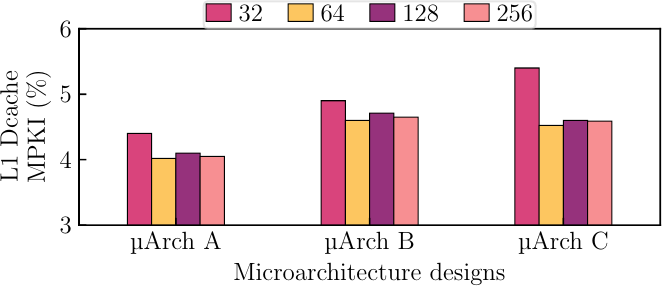}
    } \hspace{.1in}
     \subfloat[Varying $N_b$-$N_q$ for branch misprediction input. ]{
        \hspace{-.1in}\includegraphics[width=.49\linewidth]{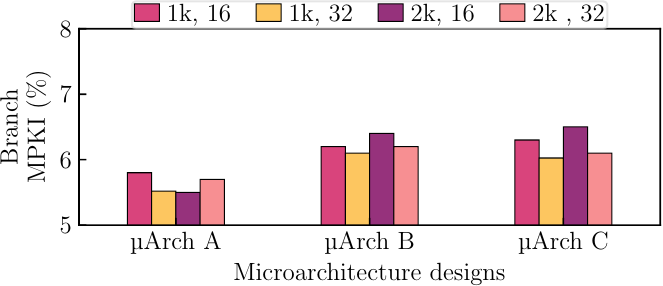}
    } 
    \caption{Choice of the context size and branch configuration.}
    \label{fig:ablation-mem-branch}
    \end{figure}

\subsection{Multi-Metric Prediction Study}
\label{sec:eval:ablation}
This section compares the prediction accuracy of L1 Dcache MPKI and branch MPKI with varying input features for their respective category. \update{We train prediction models for each parameter on training benchmarks and compare the simulation accuracy across test benchmarks to determine the best value for the parameters.}




\textbf{Data access level input.} Figure~\ref{fig:ablation-mem-branch}(a) varies the queue size of memory accesses ($N_{m}$) from 32 to 256 and evaluates average accuracy of test benchmarks for each microarchitecture design. The results indicate a general improvement in accuracy with increasing queue size for all microarchitecture configurations. However, beyond a queue size of 64, the accuracy improvements are marginal. Consequently, we select $N_{m}=64$ for generating input features related to data access level.


\textbf{Branch misprediction input.} Figure~\ref{fig:ablation-mem-branch}(b) investigates the influence of varying the combination of hash buckets $N_{b}$ (1k, 2k) and queue size $N_q$ (16,32) on the input features representing the history of branch PC address. The combination (1k,32) demonstrates favorable accuracy for predicting the branch MPKI, with no significant improvement observed by further increasing these parameters. Therefore, we opt for $N_b=1k$ and $N_q=32$ to generate the input features specifically tailored for branch instructions.

\subsection{Evaluation on Transfer Learning via Microarchitecture Agnostic Embeddings} 
\label{ref:eval:tl}

This section evaluates the effectiveness of proposed techniques for microarchitecture agnostic embeddings construction and transfer learning. 



%

\begin{figure}[h]
\centering
    \centering
    \includegraphics[width=\linewidth]{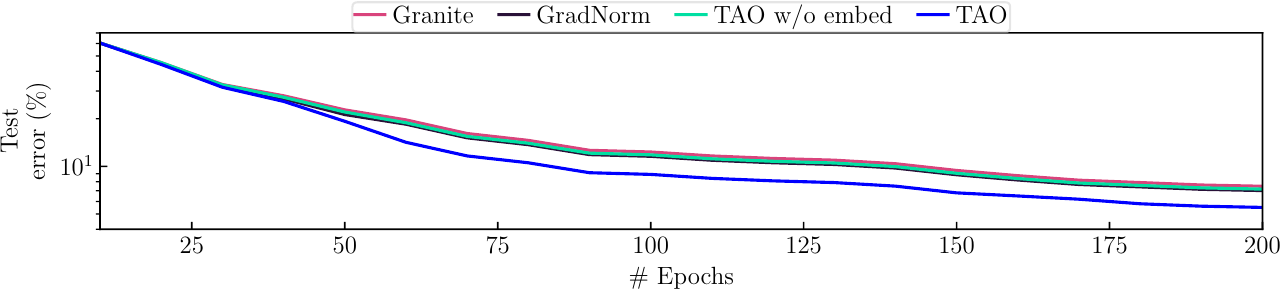}
    \caption{Number of epochs vs test error (log scale).}
    \label{fig:TL-error}
  \end{figure}

Figure~\ref{fig:TL-error} compares the test error during training of microarchitecture agnostic embedding layers for Granite, GradNorm and {\ours}. For {\ours}, we compare the performance with and without embedding adaptation layers ({\ours} {w/o embed}). The y-axis represents the average prediction error from $\mu$Arch A and $\mu$Arch B for the test dataset during the training. The x-axis represents the number of epochs for training. While training for 200 epochs, Granite and GradNorm converge with a test error of 7.5\% and 7\%, respectively. Granite exhibits the highest prediction error, attributed to its challenges in handling gradient imbalance and negative transfer. GradNorm, adept at balancing gradients from each microarchitecture prediction layer, achieves a lower error than Granite. However, it falls short of further error reduction due to negative transfer. Without using embedding adaptation layers, {\ours} demonstrates a slight improvement over Granite, achieving a test error of 7.18\% with gradient normalization.  But it falls short of surpassing GradNorm. Notably, leveraging an embedding adaptation layer with gradient normalization, {\ours} reduces prediction error further to 5.5\%. 

\begin{figure}[h]
    \centering
    \includegraphics[width=\linewidth]{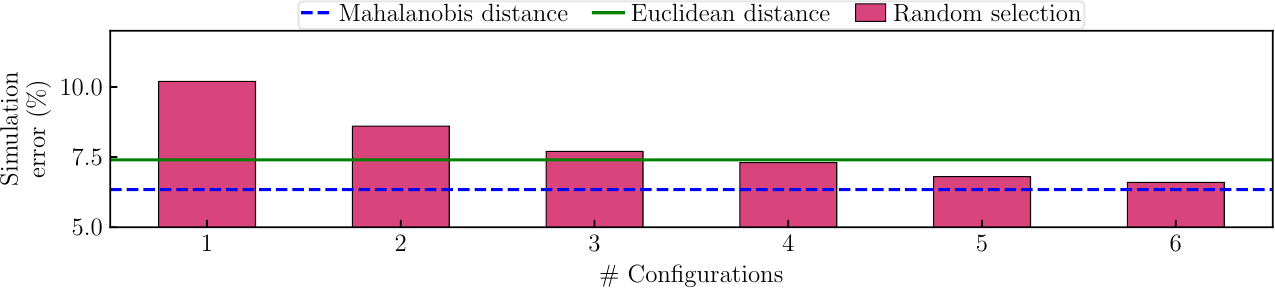}
    \caption{Training dataset selection.}
    \label{fig:model-count}  
  \end{figure}

\textbf{Training dataset.}
Figure~\ref{fig:model-count} evaluates the effectiveness of Mahalanobis distance against random selection and Euclidean distance for benchmark selection to construct microarchitecture agnostic embeddings. We prefer simulation error over training error to evaluate how well the embeddings perform with transfer learning. The y-axis represents the average simulation error for the test microarchitectures and benchmarks. We select one to six different microarchitectures randomly for random selection to construct reusable embeddings. This excludes the test $\mu$Arch A, B and C. For random selection, the simulation error starts converging after five microarchitectures.
The simulation error does not further decrease due to adversarial gradients from different microarchitectures. For Mahalanobis and Manhattan distance, we measure the performance variations of 20 designs randomly sampled from Table~\ref{table:design_space} and select two microarchitectures for training.
Euclidean distance based selection has slightly less simulation error of 7.5\% compared with 8.5\% for random selection. With Mahalanobis distance based selection, we achieve slightly better accuracy, 6.34\%, than random selection of six microarchitectures. The generated embeddings are more robust because the microarchitecture selected from Mahalanobis distance has more variations.

\textcolor{black}{\textbf{Transfer learning.} Table~\ref{table:train-time} compares the effectiveness of transfer learning to train a DL model for an unseen microarchitecture. For both {\ours} and SimNet, we train a DL model until the error during training is close to 6\%.}

\begin{wraptable}[8]{r}{8.8cm}
\update{
\begin{tabular}{|c|c|c|}
\hline
Techniques & \begin{tabular}[c]{@{}c@{}}TAO\\ (in hours)\end{tabular} & \begin{tabular}[c]{@{}c@{}}SimNet\\ (in hours)\end{tabular} \\ \hline
Scratch                         & 56  & 54 \\ \hline
Direct fine-tuning              & 38  & 41 \\ \hline
Shared embeddings + fine-tuning & 1.9 & -  \\ \hline
\end{tabular}
}
\vspace{0.01in}
\caption{
Training time.}
\label{table:train-time}
\end{wraptable}

Here, the first approach \texttt{scratch} represents a model for unseen microarchitecture trained from scratch without any transfer learning. Training a model from scratch takes 56 and 54 hours, respectively, for {\ours} and SimNet. In the second approach, \texttt{direct fine-tuning}, all parameters of the model are initialized from an earlier trained model. It takes 38 and 41 hours, respectively, for {\ours} and SimNet. The model proposed by {\ours} shows better transfer learning speed due to separated program embeddings and prediction layers. Although fine-tuning reduces the training time, it is not significant.

\update{
The third approach, \texttt{shared embeddings + fine-tuning}, is proposed by {\ours} and not directly applicable to SimNet. We use the embeddings constructed from microarchitecture agnostic embedding construction to train a model for unseen microarchitecture. The prediction layers are initialized from earlier trained models and fine-tuned. We only use 20 million instructions for fine-tuning the prediction layers. \texttt{shared embeddings + fine-tuning} further reduces the training time to only 1.9 hours. The resulting speedup comes from a reduced number of epochs for training, less inference time with shared embeddings and less per epoch time due to reduced datasets. } 



\begin{wraptable}[7]{r}{9cm}
\centering
\begin{tabular}{|cc|c|}
\hline
\multicolumn{2}{|c|}{\textbf{Training dataset selection}} &
  \multirow{2}{*}{\textbf{\begin{tabular}[c]{@{}c@{}}Training  \\ embeddings\end{tabular}}} \\ \cline{1-2}
\multicolumn{1}{|c|}{\textbf{\begin{tabular}[c]{@{}c@{}}Random design selection \\ and simulation\end{tabular}}} &
  \textbf{\begin{tabular}[c]{@{}c@{}}Dataset \\ selection\end{tabular}} &
   \\ \hline
\multicolumn{1}{|c|}{0.35 hours} &
  0.1 min &
  71 hours \\ \hline
\end{tabular}
\vspace{0.01in}
\caption{\update{Overhead of architecture agnostic embedding construction.}}
\label{table:overhead}
\end{wraptable}

\textcolor{black}{ Table~\ref{table:overhead} shows the preprocessing overhead of one-time microarchitecture agnostic embedding construction. Constructing the embeddings involves training dataset selection and training shared embeddings (refer to Section~\ref{sec:functional:embeddings}). To collect the training dataset, we first randomly sample 16 microarchitectures from the design space outlined in Table~\ref{table:design_space}, which has 184,320 total possible designs. For each sampled microarchitecture, we simulate 10 million instructions for all training benchmarks with gem5. It takes 0.35 hours to simulate and gather the performance metrics for each microarchitecture. Then, we select two microarchitectures for training the embeddings based on the Mahalanobis distance among the performance metrics of 16 microarchitectures. We use a Python script to compute the distance which takes only 0.1 min and select two microarchitectures with the maximum distance. The microarchitecture agnostic embeddings are trained from the training dataset of the two selection microarchitectures, which takes around 71 hours.}

\subsection{Hardware Design Space Exploration}
\label{sec:eval:dse}
This section aims to determine if {\ours} can be used for microarchitectural design space exploration.
For evaluation, we vary the L1 Dcache size (16KB, 32KB, 64KB, 128KB) and the branch predictors (\texttt{Local, Tournament, BiMode,} and \texttt{TAGE\_SC\_L}).


\begin{figure*}[h]
    \centering

     \subfloat[L1 Dcache MPKI.]{
        \hspace{-.1in}\includegraphics[width=.49\linewidth]{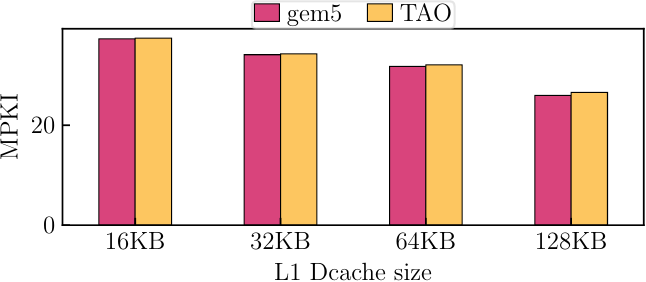}
    } \hspace{.1in}
     \subfloat[Branch MPKI.]{
        \hspace{-.1in}\includegraphics[width=.49\linewidth]{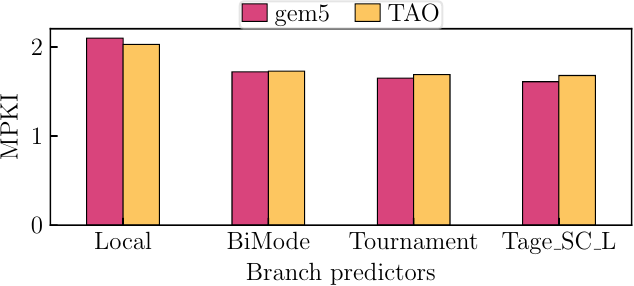}
    } 
    \caption{Hardware design space exploration of L1 Dcache misses and branch mispredictions with {\ours}. }
    \label{fig:dse}
    \end{figure*}

Figures~\ref{fig:dse}(a) compares the average cache MPKI across four test benchmarks obtained while varying L1 DCache size for gem5 simulation and {\ours}. The simulated cache MPKI decreases as the cache size increases from 16KB to 128KB. Cache MPKI predicted by {\ours} aligns with the simulated result from gem5, demonstrating that a cache size of 128KB results in the least MPKI.

In Figures~\ref{fig:dse}(b), we compare the average branch MPKI across four test benchmarks using different branch predictors for gem5 and {\ours}. The simulated result from gem5 indicates the highest branch MPKI for the \texttt{Local} and the lowest for the \texttt{Tage\_SC\_L}. Branch MPKI predicted by {\ours} also aligns with the simulated result from gem5. The prediction error is lower for simpler branch predictors like \texttt{Local}, experiencing only a marginal increase for relatively complex branch predictors like \texttt{Tage\_SC\_L}. Nonetheless, {\ours} maintains the relative accuracy across the spectrum of branch predictors. Overall, {\ours} prediction aligns with the simulated results from gem5 for hardware design exploration of L1 Dcache and branch predictors.

\section{Generality of {\ours}}

\update{
{\bf Unseen benchmarks.} {\ours} can be generalized across a wide variety of unseen benchmarks. The generality of {\ours} comes from the fact that the deep learning model is trained at the instruction level. We use multiple diverse training benchmarks to train over a variety of instructions. That allows TAO to predict performance metrics for each instruction across different benchmarks accurately. Our evaluation confirms that {\ours} maintains a good accuracy over unseen benchmarks. 

{\bf Unseen architectures.} TAO is designed to simulate single-core out-of-order superscalar processors. To simulate an unseen microarchitecture, we gather a training dataset through gem5 simulation and train the DL model with transfer learning (see Figure~\ref{fig:introduction}(d)). TAO can accommodate changes in ISAs similarly to microarchitecture changes with some additional feature engineering for ISA-specific opcodes and registers. {\ours} cannot be directly used to simulate multi-core CPU and GPU architectures. However, the techniques proposed in this paper, i.e., microarchitecture agnostic trace, embeddings, and multi-metric prediction, establish a framework for a rapid DL-based simulation and is transferable to other architectures.}

\section{Related Work}
\label{sec:related}

 Along with traditional simulation approaches, there have been parallel efforts to use ML/DL approaches for building performance models. In addition to the related work in Section~\ref{sec:intro:related}, this section discusses ML and DL-based performance models. 

\textbf{ML-based performance models.} ML-based performance models opt to build a performance model that can extrapolate the performance to unseen microarchitecture designs by simulating a few designs. Of note, these performance models aid in design exploration, different from ML-based performance models that model the runtime of applications in CPU~\cite{li2009machine,zheng2016accurate}, data centers~\cite{somashekar2024oppertune} or GPU~\cite{braun2020simple,yu2021automated,li2023path,baldini2014predicting,wu2015gpgpu}. These work use linear regression~\cite{joseph2006construction}, artificial neural network (ANN)~\cite{ipek2006efficiently}, spline functions~\cite{lee2006accurate,lee2007illustrative}, and radial basis function networks~\cite{joseph2006predictive} to build the performance model. These models take the program features and/or architecture configurations as input for predictions.
~\cite{joseph2006construction} applies linear regression to obtain estimates of performance modeled as a weighted sum of predictor variables like cache size and branch predictor.~\cite{lee2006accurate,lee2007illustrative} and ~\cite{joseph2006predictive} add spline functions and radial basis functions respectively to better model the non-linearity between the design parameters and performance.~\cite{dubach2007microarchitectural} further studies the correlation of microarchitecture parameters to the best configuration using microarchitecture specific parameters.~\cite{ipek2006efficiently} uses ANN to model the performance to automatically learn a prediction model instead of building models with domain knowledge. ~\cite{seshadri2022evaluation} proposes a GNN-based learned cost model to estimate performance metrics of ML hardware accelerators.  However, these ML-based models fail to accurately model the run-time complex dynamic interaction between the program and hardware.   
%


\textbf{DL-based detailed performance models.}
 DL-based prediction models overcome the limitations of ML-based performance models by increasing the level of abstraction at the instruction level. Ithemal~\cite{mendis2019ithemal} and Granite~\cite{sykora2022granite} are two recent works performing basic block prediction. These models first gather training datasets by collecting the basic blocks with tools like Dynamorio~\cite{bruening2012transparent}. The models predict the latency of each block separately. The input features are constructed based on each instruction and their structure in the basic blocks.  Ithemal uses LSTM to construct the embeddings for each basic block hierarchically. Meanwhile, Granite leverages the structure and dependency graph of instructions within the basic block and GNN models for throughput prediction. They are mostly used to assist compilers. Basic block throughput prediction models are limited to static basic block prediction, ignoring the impact of caches and branch prediction.

\section{Conclusion}
\label{sec:conclusion}
This paper introduces an DL-based simulator {\ours} that supports detailed, accurate and fast microarchitecture explorations. {\ours} includes three contributions: 
First, we propose a workflow for training dataset construction that allows the reuse of execution traces for simulation. Second, to increase the detail of the simulation, we redesign the input features and the DL model {using self-attention layers to support predicting a set of performance metrics of interest}. Third, we propose techniques to train a microarchitecture agnostic embedding layer {that enables fast transfer learning between different microarchitectural configurations and effectively reduces the re-training overhead of conventional DL-based simulators.} {\ours} is the first DL simulator, to the best of our knowledge, that supports detailed architecture metrics, and 18.06$\times$ faster than the state-of-the-art simulator, i.e., SimNet. 

\section{Acknowledgement}

We thank the anonymous reviewers and shepherd Derek Eager for their helpful suggestions and feedback. 
We also would like to extend our gratitude towards Cliff Young and James Laudon for their feedback on the early draft of this work. We also appreciate the support from the extended team at Google DeepMind.
This work was in part supported by the NSF CRII
Award No. 2331536, CAREER Award No. 2326141, and NSF Awards 2212370,
2319880, 2328948, 2319975 and 233130.
   
\bibliographystyle{ACM-Reference-Format}
  \bibliography{refs}


\begin{thebibliography}{79}


\ifx \showCODEN    \undefined \def \showCODEN     #1{\unskip}     \fi
\ifx \showDOI      \undefined \def \showDOI       #1{#1}\fi
\ifx \showISBNx    \undefined \def \showISBNx     #1{\unskip}     \fi
\ifx \showISBNxiii \undefined \def \showISBNxiii  #1{\unskip}     \fi
\ifx \showISSN     \undefined \def \showISSN      #1{\unskip}     \fi
\ifx \showLCCN     \undefined \def \showLCCN      #1{\unskip}     \fi
\ifx \shownote     \undefined \def \shownote      #1{#1}          \fi
\ifx \showarticletitle \undefined \def \showarticletitle #1{#1}   \fi
\ifx \showURL      \undefined \def \showURL       {\relax}        \fi
\providecommand\bibfield[2]{#2}
\providecommand\bibinfo[2]{#2}
\providecommand\natexlab[1]{#1}
\providecommand\showeprint[2][]{arXiv:#2}

\bibitem[\protect\citeauthoryear{Ahn, Li, Seongil, and Jouppi}{Ahn et~al\mbox{.}}{2013}]%
        {ahn2013mcsima+}
\bibfield{author}{\bibinfo{person}{Jung~Ho Ahn}, \bibinfo{person}{Sheng Li}, \bibinfo{person}{O Seongil}, {and} \bibinfo{person}{Norman~P Jouppi}.} \bibinfo{year}{2013}\natexlab{}.
\newblock \showarticletitle{McSimA+: A manycore simulator with application-level+ simulation and detailed microarchitecture modeling}. In \bibinfo{booktitle}{\emph{2013 IEEE International Symposium on Performance Analysis of Systems and Software (ISPASS)}}. IEEE, \bibinfo{pages}{74--85}.
\newblock


\bibitem[\protect\citeauthoryear{Akram and Sawalha}{Akram and Sawalha}{2019}]%
        {akram2019survey}
\bibfield{author}{\bibinfo{person}{Ayaz Akram} {and} \bibinfo{person}{Lina Sawalha}.} \bibinfo{year}{2019}\natexlab{}.
\newblock \showarticletitle{{A survey of computer architecture simulation techniques and tools}}.
\newblock \bibinfo{journal}{\emph{Ieee Access}}  \bibinfo{volume}{7} (\bibinfo{year}{2019}), \bibinfo{pages}{78120--78145}.
\newblock


\bibitem[\protect\citeauthoryear{Alm{\'a}si, Ca{\c{s}}caval, and Padua}{Alm{\'a}si et~al\mbox{.}}{2002}]%
        {almasi2002calculating}
\bibfield{author}{\bibinfo{person}{George Alm{\'a}si}, \bibinfo{person}{Cǎlin Ca{\c{s}}caval}, {and} \bibinfo{person}{David~A Padua}.} \bibinfo{year}{2002}\natexlab{}.
\newblock \showarticletitle{Calculating stack distances efficiently}. In \bibinfo{booktitle}{\emph{Proceedings of the 2002 workshop on Memory system performance}}. \bibinfo{pages}{37--43}.
\newblock


\bibitem[\protect\citeauthoryear{Alomar, Hamadanian, Nasr-Esfahany, Agarwal, Alizadeh, and Shah}{Alomar et~al\mbox{.}}{2023}]%
        {alomar2023causalsim}
\bibfield{author}{\bibinfo{person}{Abdullah Alomar}, \bibinfo{person}{Pouya Hamadanian}, \bibinfo{person}{Arash Nasr-Esfahany}, \bibinfo{person}{Anish Agarwal}, \bibinfo{person}{Mohammad Alizadeh}, {and} \bibinfo{person}{Devavrat Shah}.} \bibinfo{year}{2023}\natexlab{}.
\newblock \showarticletitle{CausalSim: A Causal Framework for Unbiased Trace-Driven Simulation}. In \bibinfo{booktitle}{\emph{20th USENIX Symposium on Networked Systems Design and Implementation (NSDI 23)}}. \bibinfo{pages}{1115--1147}.
\newblock


\bibitem[\protect\citeauthoryear{Arafa, Badawy, ElWazir, Barai, Eker, Chennupati, Santhi, and Eidenbenz}{Arafa et~al\mbox{.}}{2021}]%
        {arafa2021hybrid}
\bibfield{author}{\bibinfo{person}{Yehia Arafa}, \bibinfo{person}{Abdel-Hameed Badawy}, \bibinfo{person}{Ammar ElWazir}, \bibinfo{person}{Atanu Barai}, \bibinfo{person}{Ali Eker}, \bibinfo{person}{Gopinath Chennupati}, \bibinfo{person}{Nandakishore Santhi}, {and} \bibinfo{person}{Stephan Eidenbenz}.} \bibinfo{year}{2021}\natexlab{}.
\newblock \showarticletitle{{Hybrid, scalable, trace-driven performance modeling of GPGPUs}}. In \bibinfo{booktitle}{\emph{Proceedings of the International Conference for High Performance Computing, Networking, Storage and Analysis}}. \bibinfo{pages}{1--15}.
\newblock


\bibitem[\protect\citeauthoryear{Ardalani, Lestourgeon, Sankaralingam, and Zhu}{Ardalani et~al\mbox{.}}{2015}]%
        {ardalani2015cross}
\bibfield{author}{\bibinfo{person}{Newsha Ardalani}, \bibinfo{person}{Clint Lestourgeon}, \bibinfo{person}{Karthikeyan Sankaralingam}, {and} \bibinfo{person}{Xiaojin Zhu}.} \bibinfo{year}{2015}\natexlab{}.
\newblock \showarticletitle{{Cross-architecture performance prediction (XAPP) using CPU code to predict GPU performance}}. In \bibinfo{booktitle}{\emph{Proceedings of the 48th International Symposium on Microarchitecture}}. \bibinfo{pages}{725--737}.
\newblock


\bibitem[\protect\citeauthoryear{Austin, Larson, and Ernst}{Austin et~al\mbox{.}}{2002}]%
        {austin2002simplescalar}
\bibfield{author}{\bibinfo{person}{Todd Austin}, \bibinfo{person}{Eric Larson}, {and} \bibinfo{person}{Dan Ernst}.} \bibinfo{year}{2002}\natexlab{}.
\newblock \showarticletitle{{SimpleScalar: An infrastructure for computer system modeling}}.
\newblock \bibinfo{journal}{\emph{Computer}} \bibinfo{volume}{35}, \bibinfo{number}{2} (\bibinfo{year}{2002}), \bibinfo{pages}{59--67}.
\newblock


\bibitem[\protect\citeauthoryear{Bai, Huang, Wei, Ma, Li, Zheng, Yu, and Xie}{Bai et~al\mbox{.}}{2023}]%
        {bai2023archexplorer}
\bibfield{author}{\bibinfo{person}{Chen Bai}, \bibinfo{person}{Jiayi Huang}, \bibinfo{person}{Xuechao Wei}, \bibinfo{person}{Yuzhe Ma}, \bibinfo{person}{Sicheng Li}, \bibinfo{person}{Hongzhong Zheng}, \bibinfo{person}{Bei Yu}, {and} \bibinfo{person}{Yuan Xie}.} \bibinfo{year}{2023}\natexlab{}.
\newblock \showarticletitle{ArchExplorer: Microarchitecture exploration via bottleneck analysis}. In \bibinfo{booktitle}{\emph{Proceedings of the 56th Annual IEEE/ACM International Symposium on Microarchitecture}}. \bibinfo{pages}{268--282}.
\newblock


\bibitem[\protect\citeauthoryear{Baldini, Fink, and Altman}{Baldini et~al\mbox{.}}{2014}]%
        {baldini2014predicting}
\bibfield{author}{\bibinfo{person}{Ioana Baldini}, \bibinfo{person}{Stephen~J Fink}, {and} \bibinfo{person}{Erik Altman}.} \bibinfo{year}{2014}\natexlab{}.
\newblock \showarticletitle{{Predicting gpu performance from cpu runs using machine learning}}. In \bibinfo{booktitle}{\emph{2014 IEEE 26th International Symposium on Computer Architecture and High Performance Computing}}. IEEE, \bibinfo{pages}{254--261}.
\newblock


\bibitem[\protect\citeauthoryear{Bellard}{Bellard}{2005}]%
        {bellard2005qemu}
\bibfield{author}{\bibinfo{person}{Fabrice Bellard}.} \bibinfo{year}{2005}\natexlab{}.
\newblock \showarticletitle{{QEMU, a fast and portable dynamic translator.}}. In \bibinfo{booktitle}{\emph{USENIX annual technical conference, FREENIX Track}}, Vol.~\bibinfo{volume}{41}. California, USA, \bibinfo{pages}{46}.
\newblock


\bibitem[\protect\citeauthoryear{Binkert, Beckmann, Black, Reinhardt, Saidi, Basu, Hestness, Hower, Krishna, Sardashti, et~al\mbox{.}}{Binkert et~al\mbox{.}}{2011}]%
        {binkert2011gem5}
\bibfield{author}{\bibinfo{person}{Nathan Binkert}, \bibinfo{person}{Bradford Beckmann}, \bibinfo{person}{Gabriel Black}, \bibinfo{person}{Steven~K Reinhardt}, \bibinfo{person}{Ali Saidi}, \bibinfo{person}{Arkaprava Basu}, \bibinfo{person}{Joel Hestness}, \bibinfo{person}{Derek~R Hower}, \bibinfo{person}{Tushar Krishna}, \bibinfo{person}{Somayeh Sardashti}, {et~al\mbox{.}}} \bibinfo{year}{2011}\natexlab{}.
\newblock \showarticletitle{{The gem5 simulator}}.
\newblock \bibinfo{journal}{\emph{ACM SIGARCH computer architecture news}} \bibinfo{volume}{39}, \bibinfo{number}{2} (\bibinfo{year}{2011}), \bibinfo{pages}{1--7}.
\newblock


\bibitem[\protect\citeauthoryear{Brais, Kalayappan, and Panda}{Brais et~al\mbox{.}}{2020}]%
        {brais2020survey}
\bibfield{author}{\bibinfo{person}{Hadi Brais}, \bibinfo{person}{Rajshekar Kalayappan}, {and} \bibinfo{person}{Preeti~Ranjan Panda}.} \bibinfo{year}{2020}\natexlab{}.
\newblock \showarticletitle{A survey of cache simulators}.
\newblock \bibinfo{journal}{\emph{ACM Computing Surveys (CSUR)}} \bibinfo{volume}{53}, \bibinfo{number}{1} (\bibinfo{year}{2020}), \bibinfo{pages}{1--32}.
\newblock


\bibitem[\protect\citeauthoryear{Braun, Nikas, Song, Heuveline, and Fr{\"o}ning}{Braun et~al\mbox{.}}{2020}]%
        {braun2020simple}
\bibfield{author}{\bibinfo{person}{Lorenz Braun}, \bibinfo{person}{Sotirios Nikas}, \bibinfo{person}{Chen Song}, \bibinfo{person}{Vincent Heuveline}, {and} \bibinfo{person}{Holger Fr{\"o}ning}.} \bibinfo{year}{2020}\natexlab{}.
\newblock \showarticletitle{A simple model for portable and fast prediction of execution time and power consumption of GPU kernels}.
\newblock \bibinfo{journal}{\emph{ACM Transactions on Architecture and Code Optimization (TACO)}} \bibinfo{volume}{18}, \bibinfo{number}{1} (\bibinfo{year}{2020}), \bibinfo{pages}{1--25}.
\newblock


\bibitem[\protect\citeauthoryear{Bruening, Zhao, and Amarasinghe}{Bruening et~al\mbox{.}}{2012}]%
        {bruening2012transparent}
\bibfield{author}{\bibinfo{person}{Derek Bruening}, \bibinfo{person}{Qin Zhao}, {and} \bibinfo{person}{Saman Amarasinghe}.} \bibinfo{year}{2012}\natexlab{}.
\newblock \showarticletitle{{Transparent dynamic instrumentation}}. In \bibinfo{booktitle}{\emph{Proceedings of the 8th ACM SIGPLAN/SIGOPS conference on Virtual Execution Environments}}. \bibinfo{pages}{133--144}.
\newblock


\bibitem[\protect\citeauthoryear{Bucek, Lange, and v.~Kistowski}{Bucek et~al\mbox{.}}{2018}]%
        {bucek2018spec}
\bibfield{author}{\bibinfo{person}{James Bucek}, \bibinfo{person}{Klaus-Dieter Lange}, {and} \bibinfo{person}{J{\'o}akim v. Kistowski}.} \bibinfo{year}{2018}\natexlab{}.
\newblock \showarticletitle{SPEC CPU2017: Next-generation compute benchmark}. In \bibinfo{booktitle}{\emph{Companion of the 2018 ACM/SPEC International Conference on Performance Engineering}}. \bibinfo{pages}{41--42}.
\newblock


\bibitem[\protect\citeauthoryear{Butko, Garibotti, Ost, Lapotre, Gamatie, Sassatelli, and Adeniyi-Jones}{Butko et~al\mbox{.}}{2015}]%
        {butko2015trace}
\bibfield{author}{\bibinfo{person}{Anastasiia Butko}, \bibinfo{person}{Rafael Garibotti}, \bibinfo{person}{Luciano Ost}, \bibinfo{person}{Vianney Lapotre}, \bibinfo{person}{Abdoulaye Gamatie}, \bibinfo{person}{Gilles Sassatelli}, {and} \bibinfo{person}{Chris Adeniyi-Jones}.} \bibinfo{year}{2015}\natexlab{}.
\newblock \showarticletitle{{A trace-driven approach for fast and accurate simulation of manycore architectures}}. In \bibinfo{booktitle}{\emph{The 20th Asia and South Pacific Design Automation Conference}}. IEEE, \bibinfo{pages}{707--712}.
\newblock


\bibitem[\protect\citeauthoryear{CaBcaval and Padua}{CaBcaval and Padua}{2003}]%
        {cabetacaval2003estimating}
\bibfield{author}{\bibinfo{person}{Calin CaBcaval} {and} \bibinfo{person}{David~A Padua}.} \bibinfo{year}{2003}\natexlab{}.
\newblock \showarticletitle{{Estimating cache misses and locality using stack distances}}. In \bibinfo{booktitle}{\emph{Proceedings of the 17th annual international conference on Supercomputing}}. \bibinfo{pages}{150--159}.
\newblock


\bibitem[\protect\citeauthoryear{Carlson, Heirman, and Eeckhout}{Carlson et~al\mbox{.}}{2011}]%
        {carlson2011sniper}
\bibfield{author}{\bibinfo{person}{Trevor~E Carlson}, \bibinfo{person}{Wim Heirman}, {and} \bibinfo{person}{Lieven Eeckhout}.} \bibinfo{year}{2011}\natexlab{}.
\newblock \showarticletitle{{Sniper: Exploring the level of abstraction for scalable and accurate parallel multi-core simulation}}. In \bibinfo{booktitle}{\emph{Proceedings of 2011 International Conference for High Performance Computing, Networking, Storage and Analysis}}. \bibinfo{pages}{1--12}.
\newblock


\bibitem[\protect\citeauthoryear{Chen, Huang, Pandey, Li, Gao, Zheng, Ding, and Liu}{Chen et~al\mbox{.}}{2021}]%
        {chen2021re}
\bibfield{author}{\bibinfo{person}{Shiyang Chen}, \bibinfo{person}{Shaoyi Huang}, \bibinfo{person}{Santosh Pandey}, \bibinfo{person}{Bingbing Li}, \bibinfo{person}{Guang~R Gao}, \bibinfo{person}{Long Zheng}, \bibinfo{person}{Caiwen Ding}, {and} \bibinfo{person}{Hang Liu}.} \bibinfo{year}{2021}\natexlab{}.
\newblock \showarticletitle{Et: re-thinking self-attention for transformer models on gpus}. In \bibinfo{booktitle}{\emph{Proceedings of the international conference for high performance computing, networking, storage and analysis}}. \bibinfo{pages}{1--18}.
\newblock


\bibitem[\protect\citeauthoryear{Chen, Badrinarayanan, Lee, and Rabinovich}{Chen et~al\mbox{.}}{2018}]%
        {chen2018gradnorm}
\bibfield{author}{\bibinfo{person}{Zhao Chen}, \bibinfo{person}{Vijay Badrinarayanan}, \bibinfo{person}{Chen-Yu Lee}, {and} \bibinfo{person}{Andrew Rabinovich}.} \bibinfo{year}{2018}\natexlab{}.
\newblock \showarticletitle{{Gradnorm: Gradient normalization for adaptive loss balancing in deep multitask networks}}. In \bibinfo{booktitle}{\emph{International conference on machine learning}}. PMLR, \bibinfo{pages}{794--803}.
\newblock


\bibitem[\protect\citeauthoryear{Cmelik and Keppel}{Cmelik and Keppel}{1994}]%
        {cmelik1994shade}
\bibfield{author}{\bibinfo{person}{Bob Cmelik} {and} \bibinfo{person}{David Keppel}.} \bibinfo{year}{1994}\natexlab{}.
\newblock \showarticletitle{{Shade: A fast instruction-set simulator for execution profiling}}. In \bibinfo{booktitle}{\emph{Proceedings of the 1994 ACM SIGMETRICS conference on Measurement and modeling of computer systems}}. \bibinfo{pages}{128--137}.
\newblock


\bibitem[\protect\citeauthoryear{Ding and Zhong}{Ding and Zhong}{2003}]%
        {ding2003predicting}
\bibfield{author}{\bibinfo{person}{Chen Ding} {and} \bibinfo{person}{Yutao Zhong}.} \bibinfo{year}{2003}\natexlab{}.
\newblock \showarticletitle{{Predicting whole-program locality through reuse distance analysis}}. In \bibinfo{booktitle}{\emph{Proceedings of the ACM SIGPLAN 2003 conference on Programming language design and implementation}}. \bibinfo{pages}{245--257}.
\newblock


\bibitem[\protect\citeauthoryear{Dubach, Jones, and O'Boyle}{Dubach et~al\mbox{.}}{2007}]%
        {dubach2007microarchitectural}
\bibfield{author}{\bibinfo{person}{Christophe Dubach}, \bibinfo{person}{Timothy Jones}, {and} \bibinfo{person}{Michael O'Boyle}.} \bibinfo{year}{2007}\natexlab{}.
\newblock \showarticletitle{{Microarchitectural design space exploration using an architecture-centric approach}}. In \bibinfo{booktitle}{\emph{40th Annual IEEE/ACM International Symposium on Microarchitecture (MICRO 2007)}}. IEEE, \bibinfo{pages}{262--271}.
\newblock


\bibitem[\protect\citeauthoryear{Duong, Zhao, Kim, Cammarota, Valero, and Veidenbaum}{Duong et~al\mbox{.}}{2012}]%
        {duong2012improving}
\bibfield{author}{\bibinfo{person}{Nam Duong}, \bibinfo{person}{Dali Zhao}, \bibinfo{person}{Taesu Kim}, \bibinfo{person}{Rosario Cammarota}, \bibinfo{person}{Mateo Valero}, {and} \bibinfo{person}{Alexander~V Veidenbaum}.} \bibinfo{year}{2012}\natexlab{}.
\newblock \showarticletitle{{Improving cache management policies using dynamic reuse distances}}. In \bibinfo{booktitle}{\emph{2012 45Th annual IEEE/ACM international symposium on microarchitecture}}. IEEE, \bibinfo{pages}{389--400}.
\newblock


\bibitem[\protect\citeauthoryear{Elrabaa, Hroub, Mudawar, Al-Aghbari, Al-Asli, and Khayyat}{Elrabaa et~al\mbox{.}}{2017}]%
        {elrabaa2017very}
\bibfield{author}{\bibinfo{person}{Muhammad~ES Elrabaa}, \bibinfo{person}{Ayman Hroub}, \bibinfo{person}{Muhamed~F Mudawar}, \bibinfo{person}{Amran Al-Aghbari}, \bibinfo{person}{Mohammed Al-Asli}, {and} \bibinfo{person}{Ahmad Khayyat}.} \bibinfo{year}{2017}\natexlab{}.
\newblock \showarticletitle{{A very fast trace-driven simulation platform for chip-multiprocessors architectural explorations}}.
\newblock \bibinfo{journal}{\emph{IEEE Transactions on Parallel and Distributed Systems}} \bibinfo{volume}{28}, \bibinfo{number}{11} (\bibinfo{year}{2017}), \bibinfo{pages}{3033--3045}.
\newblock


\bibitem[\protect\citeauthoryear{Eyerman, Eeckhout, Karkhanis, and Smith}{Eyerman et~al\mbox{.}}{2009}]%
        {eyerman2009mechanistic}
\bibfield{author}{\bibinfo{person}{Stijn Eyerman}, \bibinfo{person}{Lieven Eeckhout}, \bibinfo{person}{Tejas Karkhanis}, {and} \bibinfo{person}{James~E Smith}.} \bibinfo{year}{2009}\natexlab{}.
\newblock \showarticletitle{{A mechanistic performance model for superscalar out-of-order processors}}.
\newblock \bibinfo{journal}{\emph{ACM Transactions on Computer Systems (TOCS)}} \bibinfo{volume}{27}, \bibinfo{number}{2} (\bibinfo{year}{2009}), \bibinfo{pages}{1--37}.
\newblock


\bibitem[\protect\citeauthoryear{Fields, Bodik, Hill, and Newburn}{Fields et~al\mbox{.}}{2003}]%
        {fields2003using}
\bibfield{author}{\bibinfo{person}{Brian~A Fields}, \bibinfo{person}{Rastislav Bodik}, \bibinfo{person}{Mark~D Hill}, {and} \bibinfo{person}{Chris~J Newburn}.} \bibinfo{year}{2003}\natexlab{}.
\newblock \showarticletitle{{Using interaction costs for microarchitectural bottleneck analysis}}. In \bibinfo{booktitle}{\emph{Proceedings. 36th Annual IEEE/ACM International Symposium on Microarchitecture, 2003. MICRO-36.}} IEEE, \bibinfo{pages}{228--239}.
\newblock


\bibitem[\protect\citeauthoryear{Goldschmidt and Hennessy}{Goldschmidt and Hennessy}{1993}]%
        {goldschmidt1993accuracy}
\bibfield{author}{\bibinfo{person}{Stephen~R Goldschmidt} {and} \bibinfo{person}{John~L Hennessy}.} \bibinfo{year}{1993}\natexlab{}.
\newblock \showarticletitle{The accuracy of trace-driven simulations of multiprocessors}.
\newblock \bibinfo{journal}{\emph{ACM SIGMETRICS Performance Evaluation Review}} \bibinfo{volume}{21}, \bibinfo{number}{1} (\bibinfo{year}{1993}), \bibinfo{pages}{146--157}.
\newblock


\bibitem[\protect\citeauthoryear{Guo, Hsueh, Leng, Qiu, Guan, Wang, Jia, Li, Guo, and Zhu}{Guo et~al\mbox{.}}{2020}]%
        {guo2020accelerating}
\bibfield{author}{\bibinfo{person}{Cong Guo}, \bibinfo{person}{Bo~Yang Hsueh}, \bibinfo{person}{Jingwen Leng}, \bibinfo{person}{Yuxian Qiu}, \bibinfo{person}{Yue Guan}, \bibinfo{person}{Zehuan Wang}, \bibinfo{person}{Xiaoying Jia}, \bibinfo{person}{Xipeng Li}, \bibinfo{person}{Minyi Guo}, {and} \bibinfo{person}{Yuhao Zhu}.} \bibinfo{year}{2020}\natexlab{}.
\newblock \showarticletitle{Accelerating sparse dnn models without hardware-support via tile-wise sparsity}. In \bibinfo{booktitle}{\emph{SC20: International Conference for High Performance Computing, Networking, Storage and Analysis}}. IEEE, \bibinfo{pages}{1--15}.
\newblock


\bibitem[\protect\citeauthoryear{Hennessy and Patterson}{Hennessy and Patterson}{2011}]%
        {hennessy2011computer}
\bibfield{author}{\bibinfo{person}{John~L Hennessy} {and} \bibinfo{person}{David~A Patterson}.} \bibinfo{year}{2011}\natexlab{}.
\newblock \bibinfo{booktitle}{\emph{{Computer architecture: a quantitative approach}} (\bibinfo{edition}{fifth} ed.)}.
\newblock \bibinfo{publisher}{Elsevier}.
\newblock


\bibitem[\protect\citeauthoryear{Hoste and Eeckhout}{Hoste and Eeckhout}{2007}]%
        {hoste2007microarchitecture}
\bibfield{author}{\bibinfo{person}{Kenneth Hoste} {and} \bibinfo{person}{Lieven Eeckhout}.} \bibinfo{year}{2007}\natexlab{}.
\newblock \showarticletitle{{Microarchitecture-independent workload characterization}}.
\newblock \bibinfo{journal}{\emph{IEEE micro}} \bibinfo{volume}{27}, \bibinfo{number}{3} (\bibinfo{year}{2007}), \bibinfo{pages}{63--72}.
\newblock


\bibitem[\protect\citeauthoryear{{\"I}pek, McKee, Caruana, de~Supinski, and Schulz}{{\"I}pek et~al\mbox{.}}{2006}]%
        {ipek2006efficiently}
\bibfield{author}{\bibinfo{person}{Engin {\"I}pek}, \bibinfo{person}{Sally~A McKee}, \bibinfo{person}{Rich Caruana}, \bibinfo{person}{Bronis~R de Supinski}, {and} \bibinfo{person}{Martin Schulz}.} \bibinfo{year}{2006}\natexlab{}.
\newblock \showarticletitle{{Efficiently exploring architectural design spaces via predictive modeling}}.
\newblock \bibinfo{journal}{\emph{ACM SIGOPS Operating Systems Review}} \bibinfo{volume}{40}, \bibinfo{number}{5} (\bibinfo{year}{2006}), \bibinfo{pages}{195--206}.
\newblock


\bibitem[\protect\citeauthoryear{Joseph, Vaswani, and Thazhuthaveetil}{Joseph et~al\mbox{.}}{2006a}]%
        {joseph2006predictive}
\bibfield{author}{\bibinfo{person}{PJ Joseph}, \bibinfo{person}{Kapil Vaswani}, {and} \bibinfo{person}{Matthew~J Thazhuthaveetil}.} \bibinfo{year}{2006}\natexlab{a}.
\newblock \showarticletitle{{A predictive performance model for superscalar processors}}. In \bibinfo{booktitle}{\emph{2006 39th Annual IEEE/ACM International Symposium on Microarchitecture (MICRO'06)}}. IEEE, \bibinfo{pages}{161--170}.
\newblock


\bibitem[\protect\citeauthoryear{Joseph, Vaswani, and Thazhuthaveetil}{Joseph et~al\mbox{.}}{2006b}]%
        {joseph2006construction}
\bibfield{author}{\bibinfo{person}{PJ Joseph}, \bibinfo{person}{Kapil Vaswani}, {and} \bibinfo{person}{Matthew~J Thazhuthaveetil}.} \bibinfo{year}{2006}\natexlab{b}.
\newblock \showarticletitle{{Construction and use of linear regression models for processor performance analysis}}. In \bibinfo{booktitle}{\emph{The Twelfth International Symposium on High-Performance Computer Architecture, 2006.}} IEEE, \bibinfo{pages}{99--108}.
\newblock


\bibitem[\protect\citeauthoryear{Kandemir, Zhao, Tang, and Karakoy}{Kandemir et~al\mbox{.}}{2015}]%
        {kandemir2015memory}
\bibfield{author}{\bibinfo{person}{Mahmut Kandemir}, \bibinfo{person}{Hui Zhao}, \bibinfo{person}{Xulong Tang}, {and} \bibinfo{person}{Mustafa Karakoy}.} \bibinfo{year}{2015}\natexlab{}.
\newblock \showarticletitle{Memory row reuse distance and its role in optimizing application performance}. In \bibinfo{booktitle}{\emph{Proceedings of the 2015 ACM SIGMETRICS International Conference on Measurement and Modeling of Computer Systems}}. \bibinfo{pages}{137--149}.
\newblock


\bibitem[\protect\citeauthoryear{Kanev, Wei, and Brooks}{Kanev et~al\mbox{.}}{2012}]%
        {kanev2012xiosim}
\bibfield{author}{\bibinfo{person}{Svilen Kanev}, \bibinfo{person}{Gu-Yeon Wei}, {and} \bibinfo{person}{David Brooks}.} \bibinfo{year}{2012}\natexlab{}.
\newblock \showarticletitle{XIOSim: power-performance modeling of mobile x86 cores}. In \bibinfo{booktitle}{\emph{Proceedings of the 2012 ACM/IEEE international symposium on Low power electronics and design}}. \bibinfo{pages}{267--272}.
\newblock


\bibitem[\protect\citeauthoryear{Khairy, Shen, Aamodt, and Rogers}{Khairy et~al\mbox{.}}{2020}]%
        {khairy2020accel}
\bibfield{author}{\bibinfo{person}{Mahmoud Khairy}, \bibinfo{person}{Zhesheng Shen}, \bibinfo{person}{Tor~M Aamodt}, {and} \bibinfo{person}{Timothy~G Rogers}.} \bibinfo{year}{2020}\natexlab{}.
\newblock \showarticletitle{{Accel-Sim: An extensible simulation framework for validated GPU modeling}}. In \bibinfo{booktitle}{\emph{2020 ACM/IEEE 47th Annual International Symposium on Computer Architecture (ISCA)}}. IEEE, \bibinfo{pages}{473--486}.
\newblock


\bibitem[\protect\citeauthoryear{Kim, Lee, Lakshminarayana, Sim, Lim, and Pho}{Kim et~al\mbox{.}}{2012}]%
        {kim2012macsim}
\bibfield{author}{\bibinfo{person}{Hyesoon Kim}, \bibinfo{person}{Jaekyu Lee}, \bibinfo{person}{Nagesh~B Lakshminarayana}, \bibinfo{person}{Jaewoong Sim}, \bibinfo{person}{Jieun Lim}, {and} \bibinfo{person}{Tri Pho}.} \bibinfo{year}{2012}\natexlab{}.
\newblock \showarticletitle{Macsim: A cpu-gpu heterogeneous simulation framework user guide}.
\newblock \bibinfo{journal}{\emph{Georgia Institute of Technology}} (\bibinfo{year}{2012}).
\newblock


\bibitem[\protect\citeauthoryear{Kim, Doppa, and Pande}{Kim et~al\mbox{.}}{2018}]%
        {kim2018machine}
\bibfield{author}{\bibinfo{person}{Ryan~Gary Kim}, \bibinfo{person}{Janardhan~Rao Doppa}, {and} \bibinfo{person}{Partha~Pratim Pande}.} \bibinfo{year}{2018}\natexlab{}.
\newblock \showarticletitle{Machine learning for design space exploration and optimization of manycore systems}. In \bibinfo{booktitle}{\emph{2018 IEEE/ACM International Conference on Computer-Aided Design (ICCAD)}}. IEEE, \bibinfo{pages}{1--6}.
\newblock


\bibitem[\protect\citeauthoryear{Krishnan, Yazdanbakhsh, Prakash, Jabbour, Uchendu, Ghosh, Boroujerdian, Richins, Tripathy, Faust, et~al\mbox{.}}{Krishnan et~al\mbox{.}}{2023}]%
        {krishnan2023archgym}
\bibfield{author}{\bibinfo{person}{Srivatsan Krishnan}, \bibinfo{person}{Amir Yazdanbakhsh}, \bibinfo{person}{Shvetank Prakash}, \bibinfo{person}{Jason Jabbour}, \bibinfo{person}{Ikechukwu Uchendu}, \bibinfo{person}{Susobhan Ghosh}, \bibinfo{person}{Behzad Boroujerdian}, \bibinfo{person}{Daniel Richins}, \bibinfo{person}{Devashree Tripathy}, \bibinfo{person}{Aleksandra Faust}, {et~al\mbox{.}}} \bibinfo{year}{2023}\natexlab{}.
\newblock \showarticletitle{Archgym: An open-source gymnasium for machine learning assisted architecture design}. In \bibinfo{booktitle}{\emph{Proceedings of the 50th Annual International Symposium on Computer Architecture}}. \bibinfo{pages}{1--16}.
\newblock


\bibitem[\protect\citeauthoryear{Larson, Chatterjee, and Austin}{Larson et~al\mbox{.}}{2001}]%
        {larson2001mase}
\bibfield{author}{\bibinfo{person}{Eric Larson}, \bibinfo{person}{Saugata Chatterjee}, {and} \bibinfo{person}{Todd~M Austin}.} \bibinfo{year}{2001}\natexlab{}.
\newblock \showarticletitle{{MASE: a novel infrastructure for detailed microarchitectural modeling.}}. In \bibinfo{booktitle}{\emph{ISPASS}}, Vol.~\bibinfo{volume}{1}. \bibinfo{pages}{9}.
\newblock


\bibitem[\protect\citeauthoryear{Lee and Brooks}{Lee and Brooks}{2006}]%
        {lee2006accurate}
\bibfield{author}{\bibinfo{person}{Benjamin~C Lee} {and} \bibinfo{person}{David~M Brooks}.} \bibinfo{year}{2006}\natexlab{}.
\newblock \showarticletitle{{Accurate and efficient regression modeling for microarchitectural performance and power prediction}}.
\newblock \bibinfo{journal}{\emph{ACM SIGOPS operating systems review}} \bibinfo{volume}{40}, \bibinfo{number}{5} (\bibinfo{year}{2006}), \bibinfo{pages}{185--194}.
\newblock


\bibitem[\protect\citeauthoryear{Lee and Brooks}{Lee and Brooks}{2007}]%
        {lee2007illustrative}
\bibfield{author}{\bibinfo{person}{Benjamin~C Lee} {and} \bibinfo{person}{David~M Brooks}.} \bibinfo{year}{2007}\natexlab{}.
\newblock \showarticletitle{{Illustrative design space studies with microarchitectural regression models}}. In \bibinfo{booktitle}{\emph{2007 IEEE 13th International Symposium on High Performance Computer Architecture}}. IEEE, \bibinfo{pages}{340--351}.
\newblock


\bibitem[\protect\citeauthoryear{Lee, Evans, and Cho}{Lee et~al\mbox{.}}{2009}]%
        {lee2009accurately}
\bibfield{author}{\bibinfo{person}{Kiyeon Lee}, \bibinfo{person}{Shayne Evans}, {and} \bibinfo{person}{Sangyeun Cho}.} \bibinfo{year}{2009}\natexlab{}.
\newblock \showarticletitle{{Accurately approximating superscalar processor performance from traces}}. In \bibinfo{booktitle}{\emph{2009 IEEE International Symposium on Performance Analysis of Systems and Software}}. IEEE, \bibinfo{pages}{238--248}.
\newblock


\bibitem[\protect\citeauthoryear{Li, Pandey, Fang, Lyv, Li, Chen, Xie, Wan, Liu, and Ding}{Li et~al\mbox{.}}{2020}]%
        {li2020ftrans}
\bibfield{author}{\bibinfo{person}{Bingbing Li}, \bibinfo{person}{Santosh Pandey}, \bibinfo{person}{Haowen Fang}, \bibinfo{person}{Yanjun Lyv}, \bibinfo{person}{Ji Li}, \bibinfo{person}{Jieyang Chen}, \bibinfo{person}{Mimi Xie}, \bibinfo{person}{Lipeng Wan}, \bibinfo{person}{Hang Liu}, {and} \bibinfo{person}{Caiwen Ding}.} \bibinfo{year}{2020}\natexlab{}.
\newblock \showarticletitle{Ftrans: energy-efficient acceleration of transformers using fpga}. In \bibinfo{booktitle}{\emph{Proceedings of the ACM/IEEE International Symposium on Low Power Electronics and Design}}. \bibinfo{pages}{175--180}.
\newblock


\bibitem[\protect\citeauthoryear{Li, Ma, Singh, Schulz, de~Supinski, and McKee}{Li et~al\mbox{.}}{2009}]%
        {li2009machine}
\bibfield{author}{\bibinfo{person}{Jiangtian Li}, \bibinfo{person}{Xiaosong Ma}, \bibinfo{person}{Karan Singh}, \bibinfo{person}{Martin Schulz}, \bibinfo{person}{Bronis~R de Supinski}, {and} \bibinfo{person}{Sally~A McKee}.} \bibinfo{year}{2009}\natexlab{}.
\newblock \showarticletitle{Machine learning based online performance prediction for runtime parallelization and task scheduling}. In \bibinfo{booktitle}{\emph{2009 IEEE international symposium on performance analysis of systems and software}}. IEEE, \bibinfo{pages}{89--100}.
\newblock


\bibitem[\protect\citeauthoryear{Li}{Li}{[n.\,d.]}]%
        {simnetgithub}
\bibfield{author}{\bibinfo{person}{Lingda Li}.} \bibinfo{year}{[n.\,d.]}\natexlab{}.
\newblock \bibinfo{title}{{Lingda-li/simnet}}.
\newblock
\newblock
\urldef\tempurl%
\url{https://github.com/lingda-li/simnet}
\showURL{%
\tempurl}


\bibitem[\protect\citeauthoryear{Li, Pandey, Flynn, Liu, Wheeler, and Hoisie}{Li et~al\mbox{.}}{2022}]%
        {simnet}
\bibfield{author}{\bibinfo{person}{Lingda Li}, \bibinfo{person}{Santosh Pandey}, \bibinfo{person}{Thomas Flynn}, \bibinfo{person}{Hang Liu}, \bibinfo{person}{Noel Wheeler}, {and} \bibinfo{person}{Adolfy Hoisie}.} \bibinfo{year}{2022}\natexlab{}.
\newblock \showarticletitle{{SimNet: Accurate and High-Performance Computer Architecture Simulation Using Deep Learning}}.
\newblock \bibinfo{journal}{\emph{Proc. ACM Meas. Anal. Comput. Syst.}} \bibinfo{volume}{6}, \bibinfo{number}{2}, Article \bibinfo{articleno}{25} (\bibinfo{date}{jun} \bibinfo{year}{2022}), \bibinfo{numpages}{24}~pages.
\newblock
\urldef\tempurl%
\url{https://doi.org/10.1145/3530891}
\showDOI{\tempurl}


\bibitem[\protect\citeauthoryear{Li, Sun, and Jog}{Li et~al\mbox{.}}{2023}]%
        {li2023path}
\bibfield{author}{\bibinfo{person}{Ying Li}, \bibinfo{person}{Yifan Sun}, {and} \bibinfo{person}{Adwait Jog}.} \bibinfo{year}{2023}\natexlab{}.
\newblock \showarticletitle{Path Forward Beyond Simulators: Fast and Accurate GPU Execution Time Prediction for DNN Workloads}. In \bibinfo{booktitle}{\emph{Proceedings of the 56th Annual IEEE/ACM International Symposium on Microarchitecture}}. \bibinfo{pages}{380--394}.
\newblock


\bibitem[\protect\citeauthoryear{Lim, Lakshminarayana, Kim, Song, Yalamanchili, and Sung}{Lim et~al\mbox{.}}{2014}]%
        {lim2014power}
\bibfield{author}{\bibinfo{person}{Jieun Lim}, \bibinfo{person}{Nagesh~B Lakshminarayana}, \bibinfo{person}{Hyesoon Kim}, \bibinfo{person}{William Song}, \bibinfo{person}{Sudhakar Yalamanchili}, {and} \bibinfo{person}{Wonyong Sung}.} \bibinfo{year}{2014}\natexlab{}.
\newblock \showarticletitle{{Power modeling for GPU architectures using McPAT}}.
\newblock \bibinfo{journal}{\emph{ACM Transactions on Design Automation of Electronic Systems (TODAES)}} \bibinfo{volume}{19}, \bibinfo{number}{3} (\bibinfo{year}{2014}), \bibinfo{pages}{1--24}.
\newblock


\bibitem[\protect\citeauthoryear{Lim and Park}{Lim and Park}{2022}]%
        {lim2022efficient}
\bibfield{author}{\bibinfo{person}{Sooyoung Lim} {and} \bibinfo{person}{Dongchul Park}.} \bibinfo{year}{2022}\natexlab{}.
\newblock \showarticletitle{Efficient Stack Distance Approximation Based on Workload Characteristics}.
\newblock \bibinfo{journal}{\emph{IEEE Access}}  \bibinfo{volume}{10} (\bibinfo{year}{2022}), \bibinfo{pages}{59792--59805}.
\newblock


\bibitem[\protect\citeauthoryear{Liu, Liang, and Gitter}{Liu et~al\mbox{.}}{2019}]%
        {liu2019loss}
\bibfield{author}{\bibinfo{person}{Shengchao Liu}, \bibinfo{person}{Yingyu Liang}, {and} \bibinfo{person}{Anthony Gitter}.} \bibinfo{year}{2019}\natexlab{}.
\newblock \showarticletitle{{Loss-balanced task weighting to reduce negative transfer in multi-task learning}}. In \bibinfo{booktitle}{\emph{Proceedings of the AAAI conference on artificial intelligence}}, Vol.~\bibinfo{volume}{33}. \bibinfo{pages}{9977--9978}.
\newblock


\bibitem[\protect\citeauthoryear{McLachlan}{McLachlan}{1999}]%
        {mclachlan1999mahalanobis}
\bibfield{author}{\bibinfo{person}{Goeffrey~J McLachlan}.} \bibinfo{year}{1999}\natexlab{}.
\newblock \showarticletitle{{Mahalanobis distance}}.
\newblock \bibinfo{journal}{\emph{Resonance}} \bibinfo{volume}{4}, \bibinfo{number}{6} (\bibinfo{year}{1999}), \bibinfo{pages}{20--26}.
\newblock


\bibitem[\protect\citeauthoryear{Mendis, Renda, Amarasinghe, and Carbin}{Mendis et~al\mbox{.}}{2019}]%
        {mendis2019ithemal}
\bibfield{author}{\bibinfo{person}{Charith Mendis}, \bibinfo{person}{Alex Renda}, \bibinfo{person}{Saman Amarasinghe}, {and} \bibinfo{person}{Michael Carbin}.} \bibinfo{year}{2019}\natexlab{}.
\newblock \showarticletitle{{Ithemal: Accurate, portable and fast basic block throughput estimation using deep neural networks}}. In \bibinfo{booktitle}{\emph{International Conference on machine learning}}. PMLR, \bibinfo{pages}{4505--4515}.
\newblock


\bibitem[\protect\citeauthoryear{Miller, Kasture, Kurian, Gruenwald, Beckmann, Celio, Eastep, and Agarwal}{Miller et~al\mbox{.}}{2010}]%
        {miller2010graphite}
\bibfield{author}{\bibinfo{person}{Jason~E Miller}, \bibinfo{person}{Harshad Kasture}, \bibinfo{person}{George Kurian}, \bibinfo{person}{Charles Gruenwald}, \bibinfo{person}{Nathan Beckmann}, \bibinfo{person}{Christopher Celio}, \bibinfo{person}{Jonathan Eastep}, {and} \bibinfo{person}{Anant Agarwal}.} \bibinfo{year}{2010}\natexlab{}.
\newblock \showarticletitle{Graphite: A distributed parallel simulator for multicores}. In \bibinfo{booktitle}{\emph{HPCA-16 2010 The Sixteenth International Symposium on High-Performance Computer Architecture}}. IEEE, \bibinfo{pages}{1--12}.
\newblock


\bibitem[\protect\citeauthoryear{Najaf-Abadi and Rotenberg}{Najaf-Abadi and Rotenberg}{2008}]%
        {najaf2008configurational}
\bibfield{author}{\bibinfo{person}{Hashem~H Najaf-Abadi} {and} \bibinfo{person}{Eric Rotenberg}.} \bibinfo{year}{2008}\natexlab{}.
\newblock \showarticletitle{{Configurational workload characterization}}. In \bibinfo{booktitle}{\emph{ISPASS 2008-IEEE International Symposium on Performance Analysis of Systems and software}}. IEEE, \bibinfo{pages}{147--156}.
\newblock


\bibitem[\protect\citeauthoryear{Ortego and Sack}{Ortego and Sack}{2004}]%
        {ortego2004sesc}
\bibfield{author}{\bibinfo{person}{Pablo~Montesinos Ortego} {and} \bibinfo{person}{Paul Sack}.} \bibinfo{year}{2004}\natexlab{}.
\newblock \showarticletitle{{SESC: SuperESCalar simulator}}. In \bibinfo{booktitle}{\emph{17 th Euro micro conference on real time systems (ECRTS’05)}}. Citeseer, \bibinfo{pages}{1--4}.
\newblock


\bibitem[\protect\citeauthoryear{Panda, Song, Dean, and John}{Panda et~al\mbox{.}}{2018}]%
        {panda2018wait}
\bibfield{author}{\bibinfo{person}{Reena Panda}, \bibinfo{person}{Shuang Song}, \bibinfo{person}{Joseph Dean}, {and} \bibinfo{person}{Lizy~K John}.} \bibinfo{year}{2018}\natexlab{}.
\newblock \showarticletitle{{Wait of a decade: Did spec cpu 2017 broaden the performance horizon?}}. In \bibinfo{booktitle}{\emph{2018 IEEE International Symposium on High Performance Computer Architecture (HPCA)}}. IEEE, \bibinfo{pages}{271--282}.
\newblock


\bibitem[\protect\citeauthoryear{Pandey, Li, Flynn, Hoisie, and Liu}{Pandey et~al\mbox{.}}{2022}]%
        {pandey2022scalable}
\bibfield{author}{\bibinfo{person}{Santosh Pandey}, \bibinfo{person}{Lingda Li}, \bibinfo{person}{Thomas Flynn}, \bibinfo{person}{Adolfy Hoisie}, {and} \bibinfo{person}{Hang Liu}.} \bibinfo{year}{2022}\natexlab{}.
\newblock \showarticletitle{Scalable deep learning-based microarchitecture simulation on GPUs}. In \bibinfo{booktitle}{\emph{SC22: International Conference for High Performance Computing, Networking, Storage and Analysis}}. IEEE, \bibinfo{pages}{1--15}.
\newblock


\bibitem[\protect\citeauthoryear{Rico, Duran, Cabarcas, Etsion, Ramirez, and Valero}{Rico et~al\mbox{.}}{2011}]%
        {rico2011trace}
\bibfield{author}{\bibinfo{person}{Alejandro Rico}, \bibinfo{person}{Alejandro Duran}, \bibinfo{person}{Felipe Cabarcas}, \bibinfo{person}{Yoav Etsion}, \bibinfo{person}{Alex Ramirez}, {and} \bibinfo{person}{Mateo Valero}.} \bibinfo{year}{2011}\natexlab{}.
\newblock \showarticletitle{{Trace-driven simulation of multithreaded applications}}. In \bibinfo{booktitle}{\emph{(IEEE ISPASS) IEEE International Symposium on Performance Analysis of Systems and Software}}. IEEE, \bibinfo{pages}{87--96}.
\newblock


\bibitem[\protect\citeauthoryear{Sanchez and Kozyrakis}{Sanchez and Kozyrakis}{2013}]%
        {sanchez2013zsim}
\bibfield{author}{\bibinfo{person}{Daniel Sanchez} {and} \bibinfo{person}{Christos Kozyrakis}.} \bibinfo{year}{2013}\natexlab{}.
\newblock \showarticletitle{ZSim: Fast and accurate microarchitectural simulation of thousand-core systems}.
\newblock \bibinfo{journal}{\emph{ACM SIGARCH Computer architecture news}} \bibinfo{volume}{41}, \bibinfo{number}{3} (\bibinfo{year}{2013}), \bibinfo{pages}{475--486}.
\newblock


\bibitem[\protect\citeauthoryear{Sener and Koltun}{Sener and Koltun}{2018}]%
        {sener2018multi}
\bibfield{author}{\bibinfo{person}{Ozan Sener} {and} \bibinfo{person}{Vladlen Koltun}.} \bibinfo{year}{2018}\natexlab{}.
\newblock \showarticletitle{{Multi-task learning as multi-objective optimization}}.
\newblock \bibinfo{journal}{\emph{Advances in neural information processing systems}}  \bibinfo{volume}{31} (\bibinfo{year}{2018}).
\newblock


\bibitem[\protect\citeauthoryear{Seshadri, Akin, Laudon, Narayanaswami, and Yazdanbakhsh}{Seshadri et~al\mbox{.}}{2022}]%
        {seshadri2022evaluation}
\bibfield{author}{\bibinfo{person}{Kiran Seshadri}, \bibinfo{person}{Berkin Akin}, \bibinfo{person}{James Laudon}, \bibinfo{person}{Ravi Narayanaswami}, {and} \bibinfo{person}{Amir Yazdanbakhsh}.} \bibinfo{year}{2022}\natexlab{}.
\newblock \showarticletitle{An evaluation of edge tpu accelerators for convolutional neural networks}. In \bibinfo{booktitle}{\emph{2022 IEEE International Symposium on Workload Characterization (IISWC)}}. IEEE, \bibinfo{pages}{79--91}.
\newblock


\bibitem[\protect\citeauthoryear{Sherwood, Perelman, Hamerly, and Calder}{Sherwood et~al\mbox{.}}{2002}]%
        {sherwood2002automatically}
\bibfield{author}{\bibinfo{person}{Timothy Sherwood}, \bibinfo{person}{Erez Perelman}, \bibinfo{person}{Greg Hamerly}, {and} \bibinfo{person}{Brad Calder}.} \bibinfo{year}{2002}\natexlab{}.
\newblock \showarticletitle{{Automatically characterizing large scale program behavior}}.
\newblock \bibinfo{journal}{\emph{ACM SIGPLAN Notices}} \bibinfo{volume}{37}, \bibinfo{number}{10} (\bibinfo{year}{2002}), \bibinfo{pages}{45--57}.
\newblock


\bibitem[\protect\citeauthoryear{Skadron, Martonosi, August, Hill, Lilja, and Pai}{Skadron et~al\mbox{.}}{2003}]%
        {skadron2003challenges}
\bibfield{author}{\bibinfo{person}{Kevin Skadron}, \bibinfo{person}{Margaret Martonosi}, \bibinfo{person}{David~I August}, \bibinfo{person}{Mark~D Hill}, \bibinfo{person}{David~J Lilja}, {and} \bibinfo{person}{Vijay~S Pai}.} \bibinfo{year}{2003}\natexlab{}.
\newblock \showarticletitle{{Challenges in computer architecture evaluation}}.
\newblock \bibinfo{journal}{\emph{Computer}} \bibinfo{volume}{36}, \bibinfo{number}{8} (\bibinfo{year}{2003}), \bibinfo{pages}{30--36}.
\newblock


\bibitem[\protect\citeauthoryear{Somashekar, Tandon, Kini, Das, Husak, Chang, Bhagwan, Natarajan, and Gandhi}{Somashekar et~al\mbox{.}}{2024}]%
        {somashekar2024oppertune}
\bibfield{author}{\bibinfo{person}{Gagan Somashekar}, \bibinfo{person}{Karan Tandon}, \bibinfo{person}{Anush Kini}, \bibinfo{person}{M Das}, \bibinfo{person}{Petr Husak}, \bibinfo{person}{CC Chang}, \bibinfo{person}{R Bhagwan}, \bibinfo{person}{N Natarajan}, {and} \bibinfo{person}{A Gandhi}.} \bibinfo{year}{2024}\natexlab{}.
\newblock \showarticletitle{Oppertune: Post-deployment configuration tuning of services made easy}. In \bibinfo{booktitle}{\emph{21st USENIX Symposium on Networked Systems Design and Implementation (NSDI 24)}}. USENIX Association.
\newblock


\bibitem[\protect\citeauthoryear{S{\`y}kora, Phothilimthana, Mendis, and Yazdanbakhsh}{S{\`y}kora et~al\mbox{.}}{2022}]%
        {sykora2022granite}
\bibfield{author}{\bibinfo{person}{Ond{\v{r}}ej S{\`y}kora}, \bibinfo{person}{Phitchaya~Mangpo Phothilimthana}, \bibinfo{person}{Charith Mendis}, {and} \bibinfo{person}{Amir Yazdanbakhsh}.} \bibinfo{year}{2022}\natexlab{}.
\newblock \showarticletitle{{GRANITE: A Graph Neural Network Model for Basic Block Throughput Estimation}}. In \bibinfo{booktitle}{\emph{2022 IEEE International Symposium on Workload Characterization (IISWC)}}. IEEE, \bibinfo{pages}{14--26}.
\newblock


\bibitem[\protect\citeauthoryear{Ubal, Jang, Mistry, Schaa, and Kaeli}{Ubal et~al\mbox{.}}{2012}]%
        {ubal2012multi2sim}
\bibfield{author}{\bibinfo{person}{Rafael Ubal}, \bibinfo{person}{Byunghyun Jang}, \bibinfo{person}{Perhaad Mistry}, \bibinfo{person}{Dana Schaa}, {and} \bibinfo{person}{David Kaeli}.} \bibinfo{year}{2012}\natexlab{}.
\newblock \showarticletitle{Multi2Sim: A simulation framework for CPU-GPU computing}. In \bibinfo{booktitle}{\emph{Proceedings of the 21st international conference on Parallel architectures and compilation techniques}}. \bibinfo{pages}{335--344}.
\newblock


\bibitem[\protect\citeauthoryear{Uhlig and Mudge}{Uhlig and Mudge}{1997}]%
        {uhlig1997trace}
\bibfield{author}{\bibinfo{person}{Richard~A Uhlig} {and} \bibinfo{person}{Trevor~N Mudge}.} \bibinfo{year}{1997}\natexlab{}.
\newblock \showarticletitle{Trace-driven memory simulation: A survey}.
\newblock \bibinfo{journal}{\emph{ACM Computing Surveys (CSUR)}} \bibinfo{volume}{29}, \bibinfo{number}{2} (\bibinfo{year}{1997}), \bibinfo{pages}{128--170}.
\newblock


\bibitem[\protect\citeauthoryear{Van~den Steen, Eyerman, De~Pestel, Mechri, Carlson, Black-Schaffer, Hagersten, and Eeckhout}{Van~den Steen et~al\mbox{.}}{2016}]%
        {van2016analytical}
\bibfield{author}{\bibinfo{person}{Sam Van~den Steen}, \bibinfo{person}{Stijn Eyerman}, \bibinfo{person}{Sander De~Pestel}, \bibinfo{person}{Moncef Mechri}, \bibinfo{person}{Trevor~E Carlson}, \bibinfo{person}{David Black-Schaffer}, \bibinfo{person}{Erik Hagersten}, {and} \bibinfo{person}{Lieven Eeckhout}.} \bibinfo{year}{2016}\natexlab{}.
\newblock \showarticletitle{{Analytical processor performance and power modeling using micro-architecture independent characteristics}}.
\newblock \bibinfo{journal}{\emph{IEEE Trans. Comput.}} \bibinfo{volume}{65}, \bibinfo{number}{12} (\bibinfo{year}{2016}), \bibinfo{pages}{3537--3551}.
\newblock


\bibitem[\protect\citeauthoryear{Wu, Greathouse, Lyashevsky, Jayasena, and Chiou}{Wu et~al\mbox{.}}{2015}]%
        {wu2015gpgpu}
\bibfield{author}{\bibinfo{person}{Gene Wu}, \bibinfo{person}{Joseph~L Greathouse}, \bibinfo{person}{Alexander Lyashevsky}, \bibinfo{person}{Nuwan Jayasena}, {and} \bibinfo{person}{Derek Chiou}.} \bibinfo{year}{2015}\natexlab{}.
\newblock \showarticletitle{GPGPU performance and power estimation using machine learning}. In \bibinfo{booktitle}{\emph{2015 IEEE 21st international symposium on high performance computer architecture (HPCA)}}. IEEE, \bibinfo{pages}{564--576}.
\newblock


\bibitem[\protect\citeauthoryear{Wunderlich, Wenisch, Falsafi, and Hoe}{Wunderlich et~al\mbox{.}}{2003}]%
        {wunderlich2003smarts}
\bibfield{author}{\bibinfo{person}{Roland~E Wunderlich}, \bibinfo{person}{Thomas~F Wenisch}, \bibinfo{person}{Babak Falsafi}, {and} \bibinfo{person}{James~C Hoe}.} \bibinfo{year}{2003}\natexlab{}.
\newblock \showarticletitle{{SMARTS: Accelerating microarchitecture simulation via rigorous statistical sampling}}. In \bibinfo{booktitle}{\emph{Proceedings of the 30th annual international symposium on Computer architecture}}. \bibinfo{pages}{84--97}.
\newblock


\bibitem[\protect\citeauthoryear{Yazdanbakhsh, Angermueller, Akin, Zhou, Jones, Hashemi, Swersky, Chatterjee, Narayanaswami, and Laudon}{Yazdanbakhsh et~al\mbox{.}}{2021}]%
        {yazdanbakhsh2021apollo}
\bibfield{author}{\bibinfo{person}{Amir Yazdanbakhsh}, \bibinfo{person}{Christof Angermueller}, \bibinfo{person}{Berkin Akin}, \bibinfo{person}{Yanqi Zhou}, \bibinfo{person}{Albin Jones}, \bibinfo{person}{Milad Hashemi}, \bibinfo{person}{Kevin Swersky}, \bibinfo{person}{Satrajit Chatterjee}, \bibinfo{person}{Ravi Narayanaswami}, {and} \bibinfo{person}{James Laudon}.} \bibinfo{year}{2021}\natexlab{}.
\newblock \showarticletitle{Apollo: Transferable architecture exploration}.
\newblock \bibinfo{journal}{\emph{arXiv preprint arXiv:2102.01723}} (\bibinfo{year}{2021}).
\newblock


\bibitem[\protect\citeauthoryear{Ye, Vijaykrishnan, Kandemir, and Irwin}{Ye et~al\mbox{.}}{2000}]%
        {ye2000design}
\bibfield{author}{\bibinfo{person}{Wu Ye}, \bibinfo{person}{Narayanan Vijaykrishnan}, \bibinfo{person}{Mahmut Kandemir}, {and} \bibinfo{person}{Mary~Jane Irwin}.} \bibinfo{year}{2000}\natexlab{}.
\newblock \showarticletitle{The design and use of simplepower: a cycle-accurate energy estimation tool}. In \bibinfo{booktitle}{\emph{Proceedings of the 37th Annual Design Automation Conference}}. \bibinfo{pages}{340--345}.
\newblock


\bibitem[\protect\citeauthoryear{Yu, Bray, Wang, Shangguan, Tang, Liu, and Chen}{Yu et~al\mbox{.}}{2021}]%
        {yu2021automated}
\bibfield{author}{\bibinfo{person}{Fuxun Yu}, \bibinfo{person}{Shawn Bray}, \bibinfo{person}{Di Wang}, \bibinfo{person}{Longfei Shangguan}, \bibinfo{person}{Xulong Tang}, \bibinfo{person}{Chenchen Liu}, {and} \bibinfo{person}{Xiang Chen}.} \bibinfo{year}{2021}\natexlab{}.
\newblock \showarticletitle{Automated runtime-aware scheduling for multi-tenant dnn inference on gpu}. In \bibinfo{booktitle}{\emph{2021 IEEE/ACM International Conference On Computer Aided Design (ICCAD)}}. IEEE, \bibinfo{pages}{1--9}.
\newblock


\bibitem[\protect\citeauthoryear{Yu, Kumar, Gupta, Levine, Hausman, and Finn}{Yu et~al\mbox{.}}{2020}]%
        {yu2020gradient}
\bibfield{author}{\bibinfo{person}{Tianhe Yu}, \bibinfo{person}{Saurabh Kumar}, \bibinfo{person}{Abhishek Gupta}, \bibinfo{person}{Sergey Levine}, \bibinfo{person}{Karol Hausman}, {and} \bibinfo{person}{Chelsea Finn}.} \bibinfo{year}{2020}\natexlab{}.
\newblock \showarticletitle{{Gradient surgery for multi-task learning}}.
\newblock \bibinfo{journal}{\emph{Advances in Neural Information Processing Systems}}  \bibinfo{volume}{33} (\bibinfo{year}{2020}), \bibinfo{pages}{5824--5836}.
\newblock


\bibitem[\protect\citeauthoryear{Zhao, Li, Shen, Liang, and Wu}{Zhao et~al\mbox{.}}{2018}]%
        {zhao2018modulation}
\bibfield{author}{\bibinfo{person}{Xiangyun Zhao}, \bibinfo{person}{Haoxiang Li}, \bibinfo{person}{Xiaohui Shen}, \bibinfo{person}{Xiaodan Liang}, {and} \bibinfo{person}{Ying Wu}.} \bibinfo{year}{2018}\natexlab{}.
\newblock \showarticletitle{{A modulation module for multi-task learning with applications in image retrieval}}. In \bibinfo{booktitle}{\emph{Proceedings of the European Conference on Computer Vision (ECCV)}}. \bibinfo{pages}{401--416}.
\newblock


\bibitem[\protect\citeauthoryear{Zheng, John, and Gerstlauer}{Zheng et~al\mbox{.}}{2016}]%
        {zheng2016accurate}
\bibfield{author}{\bibinfo{person}{Xinnian Zheng}, \bibinfo{person}{Lizy~K John}, {and} \bibinfo{person}{Andreas Gerstlauer}.} \bibinfo{year}{2016}\natexlab{}.
\newblock \showarticletitle{{Accurate phase-level cross-platform power and performance estimation}}. In \bibinfo{booktitle}{\emph{Proceedings of the 53rd Annual Design Automation Conference}}. \bibinfo{pages}{1--6}.
\newblock


\bibitem[\protect\citeauthoryear{Zhong, Shen, and Ding}{Zhong et~al\mbox{.}}{2009}]%
        {zhong2009program}
\bibfield{author}{\bibinfo{person}{Yutao Zhong}, \bibinfo{person}{Xipeng Shen}, {and} \bibinfo{person}{Chen Ding}.} \bibinfo{year}{2009}\natexlab{}.
\newblock \showarticletitle{Program locality analysis using reuse distance}.
\newblock \bibinfo{journal}{\emph{ACM Transactions on Programming Languages and Systems (TOPLAS)}} \bibinfo{volume}{31}, \bibinfo{number}{6} (\bibinfo{year}{2009}), \bibinfo{pages}{1--39}.
\newblock


\end{thebibliography}

\received{January 2023}
\received[revised]{April 2024}
\received[accepted]{April 2024}

\end{document}